\documentclass[
reprint,
superscriptaddress,
showkeys,
aps,
prd
]{revtex4-2}

\usepackage{babel,comment}
\usepackage{diagbox}
\usepackage{soul}
\usepackage{amsmath,amsfonts,mathrsfs,amssymb}
\usepackage{epsfig}
\usepackage{latexsym}
\usepackage{graphicx}
\usepackage{dcolumn}
\usepackage{bm}
\usepackage{bbm}
\usepackage[T1]{fontenc} 
\usepackage{epsf}
\usepackage{amsthm}
\usepackage{color}
\usepackage{slashed}
\usepackage{float}
\usepackage{esvect}
\usepackage{mathtools}
\usepackage{hyphenat}
\usepackage[pdfencoding=auto]{hyperref}
\newcommand{\RNum}[1]{\uppercase\expandafter{\romannumeral #1\relax}}
\DeclareMathOperator{\Tr}{Tr}

\begin{document}





\title{Holographic cold dense matter constrained by neutron stars}




\author{Lin Zhang}
\email[]{zhanglin@ucas.ac.cn}
\affiliation{School of Nuclear Science and Technology, University of Chinese Academy of Sciences,\\Beijing, P.R.China 100049}

\author{Mei Huang}
\email[]{huangmei@ucas.ac.cn}
\affiliation{School of Nuclear Science and Technology, University of Chinese Academy of Sciences,\\Beijing, P.R.China 100049}


\begin{abstract}
  The equation of state (EoS) for cold dense matter inside neutron stars is investigated by using holographic QCD models in the framework of the Einstein-Maxwell-dilaton (EMD) system and the improved Karch-Katz-Son-Stephanov (KKSS) action for matter part. This method of describing holographic nuclear matter in the EMD$+$KKSS framework is different from that by using the Dirac-Born-Infeld (DBI) action and the Chern-Simons (CS) terms. Combining with the Hebeler-Lattimer-Pethick-Schwenk (HLPS) intermediate equation of state (EoS), the hybrid EoS inside the neutron stars is constructed. The obtained hybrid EoS is located in the range that is defined by the low-density chiral effective theory, the high-density perturbative QCD, and the polytropic interpolations between them, and is constrained by the astrophysics observations. The square of the sound velocity reaches a maximum value larger than $0.8$ in the region of $2-5$ times the saturation baryon number density and approaches the conformal limit at the high baryon density range. The mass-radius relation and the tidal deformability of the neutron stars are in agreement with astrophysical measurements. The possible maximum mass for the neutron star is about $2.5 M_{\odot}$ and the radius is about $12 \mathrm{km}$ then. It is noticed that the holographic quark matter branch in the mass-radius relation is always unstable and the holographic nuclear matter can produce a stable branch. These results indicate that even in the core of the NS, the matter is still in the confinement phase and the quark matter is not favored.
\end{abstract}

\maketitle
\section{Introduction}
\label{sec_introd}

Quantum chromodynamics (QCD) is believed to be the basic theory of the strong interaction. However, the strong interaction is non-perturbative at low energy scale. Thus many phenomena such as hadron and glueball spectra, chiral phase transition and confinement-deconfinement phase transition, the equation of state (EoS) of the strong interaction matter under extreme conditions (high temperature, high baryon number density, rapid rotation, and strong magnetic field) can't be explained from the first principle until now. Therefore many efforts have been paid on developing non-perturbative methods, such as lattice QCD \cite{Kogut:1979wt,Kogut:1982ds,Ding:2015ona}, chiral effective theory \cite{Weinberg:1978kz}, Nambu--Jona-Lasinio (NJL) model \cite{Nambu:1961tp,Nambu:1961fr}, QCD sum rule \cite{Dominguez:1986td,Dominguez:1986zv,Latorre:1987wt,Narison:1988ts,Narison:1996fm,Narison:1997nw,Huang:1998wj,Narison:2008nj}, Dyson Schwinger equations (DSE) and functional renormalization group (FRG) \cite{Fu:2022gou}, and recently holographic QCD method \cite{Maldacena:1997re,Gubser:1998bc,Witten:1998qj,Kruczenski:2003uq,Sakai:2004cn,Sakai:2005yt,DaRold:2005mxj,Erlich:2005qh,Karch:2006pv}.

Exploring the equation of state (EoS) and phase structure of quantum chromodynamics (QCD) at high density is boosted both by neutron star (NS) and heavy ion collisions. Relativistic heavy ion collision (RHIC) can produce the quark-gluon plasma (QGP) with high temperature and/or high baryon number density, and strong magnetic field. Besides that, the QGP with rapid rotation can also be produced in non-central heavy ion collisions. That is the most extreme environment created in the laboratory. It has been theoretically predicted that there exists a critical end point (CEP) for QCD at finite baryon density. The search of the CEP has been one of the most important targets for the beam energy scan program at RHIC as well as for the future accelerator facilities at Facility for Antiproton and Ion Research (FAIR) in Darmstadt and Nuclotron-based Ion Collider Facility (NICA) in Dubna.

In the nature, the early universe and the compact stars are also in extreme conditions. All these provide good platforms to investigate the non-perturbative features of strong interaction and the properties of the QCD matter.

The neutron star (NS) is a kind of compact star, which is the remnant after a massive super-giant star collapses. From August 1967 on, when the existence of NSs was confirmed by the discovery of radio pulsars, more than 2700 radio pulsars have been detected. It has been wondered for more than a half\hyp{}century what's the internal structure of NSs. Whether quark matter exists inside the core of NSs \cite{Lattimer:2004pg,Annala:2019puf}? Is there any exotic matter in NSs? Do strange stars exist? We can't answer these questions yet. There are just a few of the open questions \cite{becker2009neutron}. The strong interaction matter in the core of the NS is extremely dense, the baryon number density $n_{B, \mathrm{NS}}$ of which can reach as high as $5 n_{S} - 10 n_{S}$, where $n_{S}$ is the nuclear saturation density. Although the surface of the NS is hot and the temperature can reach several hundred thousand Kelvin \cite{haensel2007neutron,kiziltan2010reassessing}, compared with the QCD energy scale $\Lambda_{\mathrm{QCD}}$ this temperature is still very low and can be neglected. Thus we can regard the matter in the NS as cold dense QCD matter.

On the experimental side, there are some observations of the NSs, including the mass measurements through the general relativistic Shapiro delay \cite{Shapiro:1964uw} and other methods \cite{Demorest:2010bx,Fonseca:2016tux,Antoniadis:2013pzd,NANOGrav:2019jur,Fonseca:2021wxt,vanKerkwijk:2010mt}, the mass and radius measurements through the detection of thermal emission from the surface of the stars \cite{Miller:2019cac,Riley:2019yda,Miller:2021qha,Riley:2021pdl}, and the radius, EoS, and tidal deformability measurements through the gravitational wave signals of the binary systems \cite{LIGOScientific:2018cki,LIGOScientific:2020zkf}. We will discuss these observations in detail in Sec. \ref{sec_NS_observ}.

Recent developments in the field of observational astronomy, e.g. the currently operating satellites, the Neutron star Interior Composition Explorer (NICER), and the gravitational-wave laser interferometers (Advanced LIGO, Virgo, and KAGRA) have provided tight constraints on neutron star masses and radii, which in turn, put strict constraints on the equation of state of dense matter.

The difficulties to predict the cold QCD matter from the theoretical side come from the awkward energy scale of the QCD matter in NS $\Lambda_{\mathrm{NS}}$: the baryon number density in the neutron star $n_{B, \mathrm{NS}}$ is high so that $\Lambda_{\mathrm{NS}}$ exceeds the appropriate range where the traditional nuclear physics theories validate; however it is still not high enough to adapt the perturbative method.

Although it's very hard to directly study the EoS of cold dense QCD matter inside the NSs, we can still derive some useful information when we combine the results of the traditional nuclear theories in the low energy scale and the results of the perturbative QCD at the high energy scale with the observation data. Under the constraints from the observation data, there are two approaches to derive the allowed range of the EoS for the cold QCD matter in the whole range of $n_{B}$: the ``direct interpolation approach'' \cite{Annala:2017llu,Annala:2021gom} and the ``causality and thermodynamic stability constraint approach'' \cite{Komoltsev:2021jzg,Gorda:2022jvk}. In the former approach, combining the observation data, a generic family of neutron-star-matter EoSs, which interpolate between theoretical results at low and high baryon density, are obtained. In the latter one, utilizing the thermodynamic stability and causality, it is found that $ab$-$initio$ calculations in QCD at high densities, as complementary to the astrophysical observations, provide significant and nontrivial information about the EoS of matter in the cores of the neutron star.

To directly calculate the NS interior composition, from the theoretical side, normally Nambu-Jona-Lasinio (NJL) model \cite{Schertler:1999xn,Lawley:2004bm,Kambe:2016olv,Malfatti:2017cln,Yazdizadeh:2019ivy,Minamikawa:2021fln,Minamikawa:2022ckn} has been used for quark matter. The holographic QCD model \cite{deBoer:2009wk,Ghoroku:2012am,Ghoroku:2013gja,Hoyos:2016zke,Jokela:2018ers,Mamani:2020pks,Ishii:2019gta,Jokela:2020piw,Jarvinen:2021jbd,Kovensky:2021ddl,Kovensky:2021kzl,Kovensky:2021wzu,Bartolini:2022rkl,Pinkanjanarod:2020mgi,Pinkanjanarod:2021qto,Burikham:2021xpn} has been developed at finite baryon density over recent years.

There are many efforts to describe the EoS of cold dense matter inside the neutron star in the holographic frame. The main approach is to use the background action in the holographic models to describe the holographic quark matter. Then the phase equilibrium conditions can be used to connect the EoS of the holographic quark matter and the EoS calculated in effective theories adapted at low energy scale. This kind of hybrid EoSs can describe the NSs. However, the problem is that the branch that corresponds to the holographic quark matter in the mass-radius relation is unstable according to the Bardeen-Thorne-Meltzer (BTM) stable criterion stated in Refs. \cite{meltzer1966normal,bardeen1966catalogue}. The BTM stable criterion is stated in the following. As one moves along the $M$-$R$ curve, where $M$ is the mass of the NS as one observes from outside and $R$ is its radius, to make sure the value of the internal energy density at the center of the star increases, all configurations remain stable until one reach the first extreme value (maximum or minimum) of $M$. The fundamental mode of radial oscillation becomes unstable at the first extreme value point. Between the first extreme value point and the second one, all configurations contain one unstable radial mode. At the second extreme value point, and also at each succeeding extreme value point, the stability of one more mode of radial oscillation changes. How the $M$-$R$ curve bends as it passes through the extreme value point determines how the stability changes. One previously stable radial mode becomes unstable at the critical point if the direction of the bend is counter-clockwise (with increasing $\epsilon_{0}$). One previously unstable mode becomes stable at the critical point if the direction of the bend is clockwise.

The key point to produce the stable branch in the holographic method is to consider the holographic nuclear matter. In Refs. \cite{Hata:2007mb,Ishii:2019gta,Jokela:2020piw,Jarvinen:2021jbd,Kovensky:2021ddl,Kovensky:2021kzl,Kovensky:2021wzu,Bartolini:2022rkl}, the Dirac-Born-Infeld (DBI) action and the Chern-Simons (CS) terms has been used to describe the baryons, where the solitons are regarded as baryons. The authors in Refs. \cite{Ishii:2019gta,Jokela:2020piw,Jarvinen:2021jbd} used the V-QCD bottom-up model action \cite{Jarvinen:2011qe,Gursoy:2007cb,Gursoy:2007er} to describe the background. In Refs. \cite{Kovensky:2021ddl,Kovensky:2021kzl,Kovensky:2021wzu}, the authors chose the Witten-Sakai-Sugimoto model \cite{Witten:1998zw,Sakai:2004cn,Sakai:2005yt} in top-down approach as the background. In Ref. \cite{Bartolini:2022rkl}, the authors used the hard-wall AdS/QCD model action to describe the background. They all obtained the stable branch that corresponds to the holographic nuclear matter.

In this work, different from the topological description of the baryons, we use the improved Karch-Katz-Son-Stephanov (KKSS) action \cite{Karch:2006pv,Fang:2016nfj,Fang:2018vkp,Fang:2018axm,Fang:2019lmd} to introduce the contribution of the hadrons and treat it as a probe. Then we can calculate the EoS of the holographic nuclear matter. This article is organized in the following. In Sec. \ref{sec_NS_observ}, we briefly summarize the observation data of the mass and radius of the NSs. In Sec. \ref{sec_HQCD_backg}, we first introduce three holographic models. Then, we solve the Einstein-Maxwell-dilaton (EMD) system to derive the thermodynamics of the background and plot the $T-\mu$ phase diagram for the confinement-deconfinement phase transition in the three models. We also compare the deformed background metric in the three models with that in the $AdS_5$ space\hyp{}time. In Sec. \ref{sec_EoS} we consider the thermodynamic contribution made by the KKSS action and calculate the EoS of the holographic nuclear matter. Combined with the Hebeler-Lattimer-Pethick-Schwenk (HLPS) intermediate EoS, we then construct the hybrid EoS of the cold strong interaction matter. In Sec. \ref{sec_NS}, we use the hybrid EoSs to calculate some properties of the neutron stars, such as mass-radius relation and tidal deformability. We also compare our results with the observation data. Finally, we give the conclusion and some discussions in Sec. \ref{sec_conclus_and_discus}.

\section{The measurement of the mass and radius of neutron stars}
\label{sec_NS_observ}

Here we give a summary of the astrophysical observations for NSs \cite{Ozel:2016oaf,Jarvinen:2021jbd}.

PSR J1614-2230 is a recycled millisecond pulsar with a very low inferred magnetic field, the magnetic induction intensity of which is $B=1.8\times10^8 \mathrm{G}$. It was discovered in a completed search of 56 unidentified mid-latitude EGRET $\gamma$-ray sources for pulsations using the Parkes telescope and the multibeam receiver in 2004 \cite{Hessels:2004ps}.
In 2010, the authors in Ref. \cite{Demorest:2010bx} presented radio timing observations of the binary millisecond pulsar PSR J1614-2230. The detection of the general relativistic Shapiro delay \cite{Shapiro:1964uw} implied the pulsar mass is $1.97 \pm 0.04 M_{\odot}$, which made it the heaviest known NS at that time. In 2016, the authors in Ref. \cite{Fonseca:2016tux} analyzed 24 binary radio pulsars in the North American Nanohertz Observatory for Gravitational Waves (NANOGrav) nine-year data set and made 14 significant measurements of the Shapiro delay. They presented the mass of PSR J1614-2230 is $1.928_{-0.017}^{+0.017} M_{\odot}$ at $68.3\%$ credibility.

PSR J0348+0432 is a mildly partially recycled binary pulsar with $P=39.1 \mathrm{ms}$ and low magnetic field ($B_\mathrm{surf}=3.1 \times 10^9 \mathrm{G}$), in a $2.4 \mathrm{hr}$ orbit with a low-mass companion, which was detected in 2012 in a $350\mathrm{MHz}$ Drift-scan Survey using the Robert C. Byrd Green Bank Telescope with the goal of finding new radio pulsars, especially millisecond pulsars that can be timed to high precision \cite{Lynch:2012vv}. PSR J0348+0432 is only the second NS with a precisely determined mass around $2 M_{\odot}$, after PSR J1614-2230. In 2013, the authors in Ref. \cite{Antoniadis:2013pzd} reported on radio-timing observations of the pulsar J0348+0432 and phase-resolved optical spectroscopy of its white-dwarf companion and then derived a NS mass in the range from $1.97 M_{\odot}$ to $2.05 M_{\odot}$ at $68.27\%$ confidence or $1.90 M_{\odot}$ to $2.18 M_{\odot}$ at $99.73\%$ confidence.

In 2014, 67 new radio pulsars including J0741+66 ($P=0.00288570374(45) \mathrm{s}$) were discovered in the Green Bank Northern Celestial Cap (GBNCC) survey, which surveys the entire sky visible to the Robert C. Byrd Green Bank Radio Telescope (GBT) at 350 MHz for radio pulsars and fast radio transients \cite{Stovall:2014gua}. In 2018, the pulsar J0741+66 was renamed J0740+6620 in Ref. \cite{Lynch:2018zxo}. In 2019, by combining data from the North American Nanohertz Observatory for Gravitational Waves (NANOGrav) 12.5-yr data set with orbital phase-specific observations using the Green Bank Telescope, the authors in Ref. \cite{NANOGrav:2019jur} measured the mass of the millisecond pulsars (MSP) J0740+6620 to be $2.14_{-0.09}^{+0.10} M_{\odot}$ at $68.3\%$ credibility interval or $2.14_{-0.18}^{+0.20} M_{\odot}$ at $95.4\%$ credibility interval. In 2021, the authors in Ref. \cite{Fonseca:2021wxt} used the data set that consisted of combined pulse arrival-time measurements made with the $100 \mathrm{m}$ Green Bank Telescope and the Canadian Hydrogen Intensity Mapping Experiment telescope. They presented the model-averaged result of the mass of PSR J0740+6620 is $2.08_{-0.07}^{+0.07} M_{\odot}$ at $68.3\%$ credible interval, which is reduced by about $2.8\%$ compared to the result given in Ref. \cite{NANOGrav:2019jur}. This model-averaged value remains the largest of all other precisely measured pulsar masses determined with the Shapiro delay to date.

There is also a special class of MSPs called black widows (and redbacks), the masses of which are higher. In black-widow systems, a millisecond pulsar is accompanied by a low-mass, few $0.01 M_{\odot}$ companion, which is bloated and strongly irradiated by the pulsar, leading to outflows strong enough to eclipse the pulsar signal for significant fractions of the orbit \cite{vanKerkwijk:2010mt}. The authors in Ref. \cite{vanKerkwijk:2010mt} present evidence that the black-widow pulsar, PSR B1957+20, has a high mass. Their best-fit pulsar mass is $2.40 \pm 0.12 M_{\odot}$. However, as it was stated in Ref. \cite{Ozel:2016oaf}, there are many difficulties to obtain accurate measurements from these ablated companions, and the possibility of even more massive NSs than J0348+0432 is not yet as robust as the results from radio timing.

These mass measurements provide a strict restriction on the EoS of the cold QCD matter, which can't be too ``soft'' and should lead to the maximum mass of NSs is about $2$ times the solar mass. Otherwise, the high masses of the above NSs cannot be explained. Many soft EoSs have already been ruled out through these mass measurements.

The measurements of the radii of the NSs are more difficult than that of the masses. In the past decade, several different techniques have been employed in the NS radius measurements. Almost all of the methods that are currently used base on the detection of thermal emission from the surface of the stars. People either measure its apparent angular size or detect the effects of the NS space\hyp{}time on this emission to extract the information of the radius. These methods can be divided into two kinds: spectroscopic measurements and timing measurements. The ongoing NASA's NS Interior Composition Explorer Mission (NICER) \cite{2012SPIE.8443E....T} uses the latter method (pulse profile modeling) where a large number of counts from a small number of rotation-powered pulsars which emit thermally from their polar caps are collected. Combined with the refined theoretical models that have been developed to analyze these waveforms, those data will provide the measurement of radii and EoSs of the NSs. The results have been published in Refs. \cite{Miller:2019cac,Riley:2019yda,Miller:2021qha,Riley:2021pdl}, in which they also use the data from the XMM-Newton telescope.

In 2019, the authors in Ref. \cite{Miller:2019cac} derived that for the isolated $205.53 \mathrm{Hz}$ millisecond pulsar PSR J0030+0451, the estimated equatorial circumferential radius is $R_{\mathrm{e}}=13.02_{-1.06}^{+1.24} \mathrm{km}$ and the gravitational mass is $M=1.44_{-0.14}^{+0.15} M_{\odot}$ at $68\%$ credible level.
In the same year, another group presented different values for PSR J0030+0451, inferred mass $M$ and equatorial radius $R_{\mathrm{eq}}$ of which are $1.34_{-0.16}^{+0.15} M_{\odot}$ and $12.71_{-1.19}^{+1.14} \mathrm{km}$, respectively, where the credible interval bounds are approximately the $16\%$ and $84\%$ quantiles in marginal posterior mass, given relative to the median \cite{Riley:2019yda}.

In 2021, the authors in Ref. \cite{Miller:2021qha} found that the equatorial circumferential radius of PSR J0740+6620 is $13.7_{-1.5}^{+2.6} \mathrm{km}$ at $68\%$ credibility based on fits of rotating hot spot patterns to NS Interior Composition Explorer (NICER) and X-ray Multi-Mirror (XMM-Newton) X-ray observations.
On the same day, another group constrained the equatorial radius and mass of PSR J0740+6620 to be $12.39_{-0.98}^{+1.30} \mathrm{km}$ and $2.072_{-0.066}^{+0.067} M_{\odot}$, respectively, each reported as the posterior credible interval bounded by the $16\%$ and $84\%$ quantiles \cite{Riley:2021pdl}.

There is another way to study the NSs, in which people observe the binary merger events involving NSs. On August 17, 2017, the Advanced LIGO and Advanced Virgo gravitational-wave detectors made their first observation of a binary NS inspiral \cite{LIGOScientific:2017vwq}. The signal, GW170817, was detected. Thereafter the multi-messenger observations of this binary NS merger are finished by telescopes and observatories ranging basically over the whole spectrum of electromagnetic waves \cite{LIGOScientific:2017ync}. In 2018, under the hypothesis that both bodies
in the GW170817 event are NSs that are described by the same equation of state and have spins within the range observed in Galactic binary NSs, the authors in Refs. \cite{LIGOScientific:2018cki} employed two methods to give an analysis of the radii and EoS of the NSs. From the LIGO and Virgo data alone and using equation-of-state-insensitive relations between various macroscopic properties of the NSs, they measured the two NS radii as $R_1=10.8_{-1.7}^{+2.0} \mathrm{km}$ for the heavier one and $R_2=10.7_{-1.5}^{+2.1} \mathrm{km}$ for the lighter one at the $90\%$ credible level. Using an efficient parametrization of the defining function $P(\epsilon)$ of the equation of state itself and requiring that the equation of state supports NSs with masses larger than $1.97 M_{\odot}$, they further derived $R_1=11.9_{-1.4}^{+1.4} \mathrm{km}$ for the heavier one and $R_2=11.9_{-1.4}^{+1.4} \mathrm{km}$ for the lighter one at the $90\%$ credible level. They also obtained constraints on $P(\epsilon)$ at supranuclear densities, with pressure at twice nuclear saturation density measured at $3.5_{-1.7}^{+2.7} \times 10^{34} \, \mathrm{dyn}\,{\mathrm{cm}}^{-2}$ at the $90\%$ level. The tidal deformability of a $1.4 M_{\odot}$\hyp{}NS could be estimated to be $\Lambda_{1.4}=190_{-120}^{+390}$ at the $90\%$ level when a common but unknown EoS is imposed. This measurement value of $\Lambda_{1.4}$ is very interesting: contrary to the constraint from the mass measurement, it suggests that ``soft'' EoSs such as APR4 are favored over ``stiff'' EoSs such as H4 or MS1. So for an EoS that satisfies both the mass measurement and the tidal deformability measurement, it should be stiff but not too stiff.

First identified in data from two detectors, LIGO Livingston and Virgo on 2019 August 14, GW190814 is the signal of a compact binary coalescence with the most unequal mass ratio yet measured with gravitational waves: $q=0.112_{-0.009}^{+0.008}$ \cite{LIGOScientific:2020zkf}. The primary component of GW190814 is conclusively a black\hyp{}hole (BH) with mass $m_1=23.2_{-1.0}^{+1.1} M_{\odot}$. Due to the lack of measurable tidal deformations and the absence of an electromagnetic counterpart, the secondary component may be an NS or a BH. However, comparing with the maximum NS mass predicted by studies of GW170817's remnant indicate it's an NS by current knowledge of the NS equation of state, and by electromagnetic observations of NSs in binary systems \cite{LIGOScientific:2020zkf}. If it's true, this source with the mass of $2.59_{-0.09}^{+0.08} M_{\odot}$ is the heaviest NS ever observed in a double compact-object system. Then the corresponding constraints on the radius and tidal deformability of a canonical $1.4 M_{\odot}$\hyp{}NS are $R_{1.4}=12.9_{-0.7}^{+0.8} \mathrm{km}$ and $\Lambda_{1.4}=616_{-158}^{+273}$ \cite{LIGOScientific:2020zkf}.

Finally, we conclude all the data in this section in Tab. \ref{tab_NS_mass_radius_index_1} and Tab. \ref{tab_NS_mass_radius_index_2}.
\begin{table}[htbp]
  \begin{small}
    \begin{tabular}{|c|c|c|}
      \hline
      pulsar     & mass $M$ ($M_{\odot}$)                           & radius $R$ ($\mathrm{km}$)                    \\
      \hline
      J1614-2230 & $1.97 \pm 0.04$ \cite{Shapiro:1964uw}            & --                                            \\
                 & $1.928_{-0.017}^{+0.017}$ \cite{Fonseca:2016tux} &                                               \\
      \hline
      J0348+0432 & $1.97-2.05$ \cite{Antoniadis:2013pzd}            & --                                            \\
      \hline
      J0740+6620 & $2.14_{-0.09}^{+0.10}$ \cite{NANOGrav:2019jur}   & --                                            \\
                 & $2.08_{-0.07}^{+0.07}$ \cite{Fonseca:2021wxt}    &                                               \\
      \hline
      B1957+20   & $2.40 \pm 0.12$ \cite{vanKerkwijk:2010mt}        & --                                            \\
      \hline
      J0030+0451 & $1.44_{-0.14}^{+0.15}$ \cite{Miller:2019cac}     & $13.02_{-1.06}^{+1.24}$ \cite{Miller:2019cac} \\
                 & $1.34_{-0.16}^{+0.15}$ \cite{Riley:2019yda}      & $12.71_{-1.19}^{+1.14}$ \cite{Riley:2019yda}  \\
      \hline
      J0740+6620 &                                                  & $13.7_{-1.5}^{+2.6}$ \cite{Miller:2021qha}    \\
                 & $2.072_{-0.066}^{+0.067}$ \cite{Riley:2021pdl}   & $12.39_{-0.98}^{+1.30}$ \cite{Riley:2021pdl}  \\
      \hline
    \end{tabular}
    \caption{The mass and radius for several discovered pulsars.}
    \label{tab_NS_mass_radius_index_1}
  \end{small}
\end{table}

\begin{table*}[htbp]
  \begin{small}
    \begin{tabular}{|c|c|c|c|c|c|c|}
      \hline
      gravitational-wave signal & mass $M$ ($M_{\odot}$)                                                             & radius $R$ ($\mathrm{km}$)                                                       & tidal deformability $\Lambda_{1.4}$               \\
      \hline
      GW170817                  & --                                                                                 & $R_1=10.8_{-1.7}^{+2.0}$, $R_2=10.7_{-1.5}^{+2.1}$ \cite{LIGOScientific:2018cki} & $190_{-120}^{+390}$ \cite{LIGOScientific:2018cki} \\
                                &                                                                                    & $R_1=11.9_{-1.4}^{+1.4}$, $R_2=11.9_{-1.4}^{+1.4}$ \cite{LIGOScientific:2018cki} &                                                   \\
      \hline
      GW190814                  & $m_1=23.2_{-1.0}^{+1.1}$, $m_2=2.59_{-0.09}^{+0.08}$ \cite{LIGOScientific:2020zkf} & --                                                                               & $616_{-158}^{+273}$ \cite{LIGOScientific:2020zkf} \\
      \hline
    \end{tabular}
    \caption{Some discovery parameters and derived parameters for two binary compact object systems from the gravitational-wave detection.}
    \label{tab_NS_mass_radius_index_2}
  \end{small}
\end{table*}

\section{The holographic framework and the background of the holographic models}
\label{sec_HQCD_backg}

Firstly we introduce the framework of the Einstein-Maxwell-dilaton (EMD) system, which comes back to the gravity-dilaton coupling system at zero chemical potential. As the background action, the EMD system is widely used in the holographic QCD models. It can be used to describe the QCD confinement-deconfinement phase diagram \cite{Cai:2012xh,Chen:2019rez}, the effect of the magnetic field on the QCD phase transition \cite{He:2020fdi}, the rotating effect on deconfinement phase transition \cite{Chen:2020ath}, the thermodynamic properties of QCD and the location of the critical endpoint (CEP) \cite{Cai:2022omk}, the critical behavior near the CEP \cite{Li:2017ple,Chen:2018vty}, and the spectra glueballs and corresponding EoS \cite{Zhang:2021itx}.

Including hadrons excitation, the total action of the $5$-dimensional holographic QCD model is
\begin{align}
   & S_{\text {total }}^{s}=S_{b}^{s}+S_{m}^{s},
  \label{action_total}
\end{align}
where $S_{b}^{s}$ is the action for the background in the string frame, and $S_{m}^{s}$ is the action that describes hadrons in the string frame.


The EMD action in the string frame is denoted by $S_{b}^{s}$:
\begin{alignat}{3}
   & S_{b}^{s}=\frac{1}{2 \kappa_{5}^{2}} & \int & \mathrm{d}^{5} x \sqrt{-g^{s}} e^{-2 \Phi} \Big[ R^{s}+4 {g^{s}}^{M N} \partial_{M} \Phi \partial_{N} \Phi
  \nonumber                                                                                                                                                                                               \\
   &                                      &      & -V^{s}(\Phi)-\frac{h(\Phi)}{4} e^{\frac{4 \Phi}{3}} {g^{s}}^{M \widetilde{M} }{g^{s}}^{N \widetilde{N} } F_{M N} F_{\widetilde{M} \widetilde{N}}\Big],
  \label{action_EMD_11}
\end{alignat}
where $s$ denotes the string frame, $\kappa_{5}^{2}=8 \pi G_5$, $G_5$ is the $5$-dimensional Newton constant. $g^s$ is the determinant of the metric in the string frame: $g^s=\det \left(g_{M N }\right)$. The metric tensor in the string frame can be extracted from
\begin{align}
   & d s^{2}=\frac{L^2 e^{2 A_{s}(z)}}{z^{2}}\bigg(-f(z) d t^{2}+\frac{d z^{2}}{f(z)}+d y_{1}^{2}
  \nonumber                                                                                       \\
   & \qquad \qquad \qquad \qquad +d y_{3}^{2}+d y_{3}^{2}\bigg),
  \label{metric_str}
\end{align}
where $L$ is the curvature radius of the asymptotic $AdS_5$ space\hyp{}time. For simplicity and without loss of generality, we assume $L=1$ in the following calculations.
$R^s$ is the Ricci curvature scalar in the string frame. The $5$-dimensional scalar field $\Phi(z)$ is the dilaton field that depends only on the coordinate $z$. $F_{M N}$ is the field strength of the $U(1)$ gauge field $A_{M}$:
\begin{align}
  F_{M N}=\partial_{M} A_{N}-\partial_{N} A_{M}.
  \label{field_strength_1}
\end{align}
The $5$-dimensional vector field $A_{M}$ is dual to the baryon number current. The function $h(\Phi)$ describes the coupling strength of $A_{M}$ in the theory, and the function $V^{s}(\Phi)$ represents the potential of the dilaton field in the string frame. $h(\Phi)$ and $V^{s}(\Phi)$ are the functions that depend only on the value of $\Phi$.

\subsection{The Einstein-Maxwell-dilaton system}
\label{subsec_EMD}
As discussed in Ref. \cite{Li:2011hp}, it is more convenient to choose the string frame when solving the vacuum expectation value of the loop operator, and it is more convenient to choose the Einstein frame to work out the gravity solution and to study the equation of state. So we apply the following Weyl transformation \cite{weyl1921raum,weyl1993space} to Eq. (\ref{action_EMD_11}):
\begin{align}
  g_{M N}^{s}=\mathrm{e}^{\frac{4}{3}\Phi}  g_{M N}^{E},
  \label{Weyl_transf}
\end{align}
where $g_{M N}^{E}$ is the metric tensor in the Einstein frame. The capital letter 'E' denotes the Einstein frame.
Then, Eq. (\ref{action_EMD_11}) becomes
\begin{alignat}{3}
   & S^E =\frac{1}{2 \kappa_{5}^{2}} & \int & \mathrm{d}^{5} x \sqrt{-g^E} \Big[ R^E-\frac{4}{3} {g^E}^{M N} \partial_{M} \Phi \partial_{N} \Phi
  \nonumber                                                                                                                                                                   \\
   &                                 &      & -V^E(\Phi)-\frac{h(\Phi)}{4} {g^{E}}^{M \widetilde{M} }{g^{E}}^{N \widetilde{N} } F_{M N} F_{\widetilde{M} \widetilde{N}}\Big],
  \label{action_EMD_2}
\end{alignat}
where the function $V^E=\mathrm{e}^{\frac{4}{3}\Phi} V^{s}$.

The coefficient of the kinetic term of the dilaton field $\Phi$ is $\frac{4}{3}$ in Eq. (\ref{action_EMD_2}), which is not canonical. So we define a new dilaton field $\phi$:
\begin{align}
   & \phi=\sqrt{\frac{8}{3}}\Phi.
  \label{new_dilaton}
\end{align}
Now the action Eq. (\ref{action_EMD_2}) becomes
\begin{align}
   & S^E =\int \mathrm{d}^{5} {\mathcal{L}}^E,
  \label{action_EMD_3}
\end{align}
where ${\mathcal{L}}^E$ is the Lagrangian density in the Einstein frame:
\begin{alignat}{3}
   & {\mathcal{L}}^E =\frac{1}{2 \kappa_{5}^{2}} &  & \sqrt{-g^E} \Big[ R^E-\frac{1}{2} {g^E}^{M N} \left( \partial_{M} \phi \right) \left( \partial_{N} \phi \right)
  \nonumber                                                                                                                                                                                       \\
   &                                             &  & -V_{\phi}(\phi)-\frac{h_{\phi}(\phi)}{4} {g^{E}}^{M \widetilde{M} }{g^{E}}^{N \widetilde{N} } F_{M N} F_{\widetilde{M} \widetilde{N}}\Big].
  \label{Lagrangian_EMD_1}
\end{alignat}
The function $V_\phi(\phi)=V^E(\Phi)$, and $h_{\phi}(\phi)=h(\Phi)$. According to Eqs. (\ref{metric_str}), (\ref{Weyl_transf}) and (\ref{new_dilaton}), we then derive the line element of the space\hyp{}time in the Einstein frame:
\begin{align}
   & d s^{2}=\frac{L^2 e^{2 A_{E}(z)}}{z^{2}}\bigg(-f(z) d t^{2}+\frac{d z^{2}}{f(z)}+d y_{1}^{2}
  \nonumber                                                                                       \\
   & \qquad \qquad \qquad \qquad +d y_{3}^{2}+d y_{3}^{2}\bigg),
  \label{metric_Einstein}
\end{align}
where
\begin{align}
   & A_E(z)=A_s(z)-\sqrt{\frac{1}{6}}\phi(z).
  \label{relat_of_A}
\end{align}
Using the variation method, we can derive the Einstein field equations and the equations of motion of $A_{M}$ and $\phi$ from Eq. (\ref{action_EMD_3})-(\ref{Lagrangian_EMD_1}):
\begin{align}
   & R_{M N}^{E}-\frac{1}{2} g_{M N}^{E} R^{E}-T_{M N}=0, \nonumber                                                                                                                                    \\
   & \nabla_{M}\left[h_{\phi}(\phi) F^{M N}\right]=0, \nonumber                                                                                                                                        \\
   & \partial_{M}\left[\sqrt{-g} \partial^{M} \phi\right]-\sqrt{-g}\left(\frac{\mathrm{d} V_{\phi}(\phi)}{\mathrm{d} \phi}+\frac{F^{2}}{4} \frac{\mathrm{d} h_{\phi}(\phi)}{\mathrm{d} \phi}\right)=0,
  \label{EoMs_1}
\end{align}
where $T_{M N}$ is the energy-momentum tensor:
\begin{alignat}{3}
   & T_{M N} & = & \frac{1}{2}\bigg[\left(\partial_{M} \phi \right) \left(\partial_{N} \phi \right) -\frac{1}{2} g_{M N}^{E} {g^E}^{P \widetilde{P}}\left(\partial_{P} \phi\right) \left(\partial_{\widetilde{P}} \phi\right)
  \nonumber                                                                                                                                                                                                                   \\
   &         &   & -g_{M N}^{E} V_{\phi}(\phi)\bigg] +\frac{h_{\phi}(\phi)}{2}\bigg({g^E}^{P \widetilde{P}} F_{M P} F_{N \widetilde{P}}
  \nonumber                                                                                                                                                                                                                   \\
   &         &   & -\frac{1}{4} g_{M N}^{E} {g^{E}}^{P \widetilde{P} }{g^{E}}^{Q \widetilde{Q} } F_{P Q} F_{\widetilde{P} \widetilde{Q}}\bigg).
  \label{energy_momentum_tensor_1}
\end{alignat}

It's safe to suppose all the components of the vector field $A_{M}(z)$ are zero except the $t$-component $A_{t}(z)$. After substituting the metric Eq. (\ref{metric_Einstein}) into the equations of motion (EoMs) Eq. (\ref{EoMs_1}), we then get the EoMs for the components of the fields:
\begin{align}
   & A_{t}^{\prime \prime}+A_{t}^{\prime}\left(-\frac{1}{z}+\frac{{h_{\phi}}^{\prime}}{h_{\phi}}+{A_{E}}^{\prime}\right)=0,
  \label{EoMs_2_1}                                                                                                                                                                         \\
   & f^{\prime \prime}+f^{\prime}\left(-\frac{3}{z}+3 {A_{E}}^{\prime}\right)-\frac{e^{-2 {A_{E}}} A_{t}^{\prime 2} z^{2} h_{\phi}}{L^2}=0,
  \label{EoMs_2_2}                                                                                                                                                                         \\
   & A_{E}^{\prime \prime} +\frac{f^{\prime \prime}}{6 f}+A_{E}^{\prime}\left(-\frac{6}{z}+\frac{3 f^{\prime}}{2 f}\right)-\frac{1}{z}\left(-\frac{4}{z}+\frac{3 f^{\prime}}{2 f}\right)
  \nonumber                                                                                                                                                                                \\
   & +3 {A_{E}}^{\prime 2} +\frac{L^2 e^{2 {A_{E}}} V_{\phi}}{3 z^{2} f}=0,
  \label{EoMs_2_3}                                                                                                                                                                         \\
   & A_{E}^{\prime \prime}-A_{E}^{\prime}\left(-\frac{2}{z}+A_{E}^{\prime}\right)+\frac{\phi^{\prime 2}}{6}=0,
  \label{EoMs_2_4}                                                                                                                                                                         \\
   & \phi^{\prime \prime}+\phi^{\prime}\left(-\frac{3}{z}+\frac{f^{\prime}}{f}+3 A_{E}^{\prime}\right)-\frac{L^2 e^{2 {A_{E}}}}{z^{2} f} \frac{\mathrm{d} V_{\phi}(\phi)}{\mathrm{d} \phi}
  \nonumber                                                                                                                                                                                \\
   & +\frac{z^{2} e^{-2 {A_{E}}} A_{t}^{\prime 2}}{2 L^2 f} \frac{\mathrm{d} h_{\phi}(\phi)}{\mathrm{d} \phi}=0.
  \label{EoMs_2_5}
\end{align}
Only 4 equations are independent in the above 5 equations. So we choose Eq. (\ref{EoMs_2_5}) as a constraint and use it to check the solutions of the EoMs.

\subsection{The matter part action in the holographic model}
\label{subsec_matter_part}

As mentioned earlier, $S_{m}^{s}$ in Eq. (\ref{action_total}) is the action that describe hadrons in the string frame. Here we use the improved KKSS action:

\begin{align}
   & S_{m}^{s}=-\beta \int \mathrm{d}^5 x  \sqrt{-g^s} \mathrm{e}^{-\Phi} \Tr \Big\{ \left| D_M X \right|^2 + V_X^s(X_M)
  \nonumber                                                                                                              \\
   & \qquad \qquad \qquad \qquad \qquad \qquad \quad+\frac{1}{4 g_5^2} \left( F_L^2+F_R^2 \right) \Big\}
  \nonumber                                                                                                              \\
   & \quad \qquad+S_{\text{baryons}}^{s},
  \label{action_matter}
\end{align}
where
\begin{align}
  D_M X=\nabla_M X-\mathrm{i} g_c A_M X
  \label{covar_der}
\end{align}
is the covariant derivative of the $5$-dimensional scalar field $X$, \begin{align}
   & V_X^s(X_M)={M_X}_5^2 \left| X \right|^2+\lambda_4 \left| X \right|^4
  \label{potent_X}
\end{align}
is the potential of the $5$-dimensional scalar field $X$. The parameter $\beta$ describe the coupling strength between the matter part and the background in the action. $D^{M} X=\partial^{M} X-\mathrm{i} A_{L}^{M} X+\mathrm{i} X A_{R}^{M}$, $A_{L}^{M}=A_{L}^{a, M} t_{L}^{a}$, $A_{R}^{M}=A_{R}^{a, M} t_{R}^{a}$, $t_{L}^{a}$ and $t_{R}^{a}$ are the generators of the $SU(2)_L$ and $SU(2)_R$ group, respectively. $g_5^2=\frac{12 \pi^2}{N_c}=4 \pi^2$. $S_{\text{baryons}}$ describes the baryons. The original KKSS action is proposed in Ref. \cite{Karch:2006pv}. It is modified in Refs. \cite{Fang:2016nfj,Fang:2018vkp,Fang:2018axm,Fang:2019lmd} to describe the chiral phase transition and meson spectra.

According to the AdS/CFT dictionary \cite{Erlich:2005qh}, the bulk scalar field $X$ and the chiral gauge fields $A_{L,R}^M$ are dual to the relevant QCD operators at the ultraviolet (UV) boundary $z=0$. The bulk scalar field $X$ can be decomposed as
\begin{align}
   & X=\left(\frac{\chi (z)}{2}+S(x,z)\right) \mathrm{e}^{2 \mathrm{i} \pi(x,z)},
  \label{decomp_X}
\end{align}
where $\pi(x,z)=\pi^a(x,z) t^a$ is the pseudoscalar meson field and $S(x,z)$ is the scalar meson field. The field $\chi(z)$ is related to the vacuum expectation value (VEV):
\begin{align}
  \langle X\rangle=\frac{\chi}{2} I_{2},
  \label{VEV_X}
\end{align}
where $I_2$ is the $2\times 2$ identity matrix. The $5$-dimensional bulk gauge field $A_{L,R}^M$ can be recombined into the vector field $V^M$ and the axial-vector field $A^M$:
\begin{align}
   & V^M=\frac{1}{2}(A_L^M+A_R^M),
   & A^M=\frac{1}{2}(A_L^M-A_R^M).
  \label{vec_and_axial_vec_filed}
\end{align}
The field-strength tensor for the vector field and the axial-vector field are
\begin{align}
   & F_{V}^{M N} =\frac{1}{2}\left(F_{L}^{M N}+F_{R}^{M N}\right)
  \nonumber                                                                                                                 \\
   & \qquad =\partial^{M} V^{N}-\partial^{N} V^{M}-\mathrm{i}\left[V^{M}, V^{N}\right]-\mathrm{i}\left[A^{M}, A^{N}\right],
  \\
   & F_{A}^{M N} =\frac{1}{2}\left(F_{L}^{M N}-F_{R}^{M N}\right)
  \nonumber                                                                                                                 \\
   & \qquad =\partial^{M} A^{N}-\partial^{N} A^{M}-\mathrm{i}\left[V^{M}, A^{N}\right]-\mathrm{i}\left[A^{M}, V^{N}\right].
  \label{vec_and_axial_vec_filed_strength}
\end{align}

According to mass-dimension relationship $M^2=(\Delta-p)(\Delta+p-4)$ and $\Delta=3$, $p=0$, the $5$-dimensional mass square of the bulk field $X$ is $M_X^2=-3$.

The action $S_{m}^{s}$ that describes the matter part can be decomposed as
\begin{align}
   & S_{m}^{s}=S_{\chi}^{s}+S_{\text{mesons}}^{s}+S_{\text{baryons}}^{s},
  \label{action_matter_index2}
\end{align}
where
\begin{align}
   & S_{\chi}=-\beta \int \mathrm{d}^5 x  \sqrt{-g^s} \mathrm{e}^{-\Phi} \bigg\{  \frac{1}{2} \left| \partial_M \chi  -\mathrm{i} g_c A_M \chi \right|^2
  \nonumber                                                                                                                                              \\
   & \qquad \qquad \qquad \qquad \qquad \qquad + V_{\chi}^s(\chi) \bigg\}
  \label{action_chi}
\end{align}
is the action for the VEV $\chi$. The $5$-dimensional fields that dual to the mesons and baryons tower are perturbations compared with the field $\chi$. The contribution made by $S_{\text{mesons}}^{s}$ and $S_{\text{baryons}}^{s}$ to the thermodynamics is believed to be small compared with that from $S_{\chi}^{s}$. Thus we will neglect $S_{\text{mesons}}^{s}$ and $S_{\text{baryons}}^{s}$ in the calculation in this article.

The EoM of the field $\chi(z)$ can be obtained from the action Eq. (\ref{action_matter}):
\begin{align}
   & \chi''(z)+\left(3 {A_E}'(z)+\frac{\sqrt{6}}{4}\phi'(z)-\frac{3}{z}+\frac{f'(z)}{f(z)}\right) \chi'(z)
  \nonumber                                                                                                                  \\
   & -\frac{1}{f(z)} \frac{1}{z^2}\mathrm{e}^{2 A_E(z)+\frac{\sqrt{6}}{3}\phi(z)}
  \nonumber                                                                                                                  \\
   & \times \bigg[M_X^2 \chi(z)-g_c^2 z^2 \mathrm{e}^{-2 A_E(z)-\frac{\sqrt{6}}{3}\phi(z)} \frac{1}{f(z)} {A_t(z)}^2 \chi(z)
  \nonumber                                                                                                                  \\
   & \qquad +\frac{1}{2}\lambda_4 {\chi(z)}^3\bigg]=0.
  \label{EoM_of_chi}
\end{align}
Here we use the Eq. (\ref{new_dilaton}) and Eq. (\ref{relat_of_A}).

The asymptotic expansion of the field $\chi(z)$ in the UV region is
\begin{align}
  \chi(z\rightarrow 0)=m_q \zeta z+\frac{\sigma}{\zeta} z^3+\cdots,
  \label{UV_asymp_expans_of_chi}
\end{align}
where $m_q=3.5 \mathrm{MeV}$ \cite{PhysRevD.98.030001} is the current quark mass, $\sigma$ is the chiral condensate, $\zeta=\frac{\sqrt{N_c}}{2\pi}=\frac{\sqrt{3}}{2\pi}$ is the normalization constant.

\subsection{Three different holographic models}
\label{subsec_three_models}
In this work, we take three different holographic models.

\subsubsection{Model \texorpdfstring{\RNum{1}}{1}}
\label{subsec_model_1}

Model \RNum{1} is a variant of Gubser model \cite{DeWolfe:2010he}. The potential for the dilation field $\phi(z)$ and the function $h_{\phi}(\phi)$ are taken from Ref. \cite{Cai:2022omk}:
\begin{align}
  V_{\phi}(\phi)=-12 \cosh (c_1 \phi)+(6 c_1^2 -\frac{3}{2}){\phi}^2+c_2 {\phi}^6,
  \label{V_phi_model_1}
\end{align}
\begin{align}
  h_{\phi}(\phi)=\frac{1}{1+c_3} \text{sech} (c_4 {\phi}^3)+\frac{c_3}{1+c_3} {\mathrm{e}}^{-c_5 \phi},
  \label{h_phi_model_1}
\end{align}
where $c_1=0.71$, $c_2=0.0037$, $c_3=1.935$, $c_4=0.085$, and $c_5=30$. The $5$-dimensional Newtonian constant is $G_5=0.42$. The characteristic energy scale \cite{Critelli:2017oub} of the EMD system is $\Lambda=1.085 \mathrm{GeV}$. The Eqs. (\ref{V_phi_model_1})-(\ref{h_phi_model_1}) and the values of the above parameters are taken from Ref. \cite{Cai:2022omk}, which can give the thermodynamic properties in good agreement with the lattice simulations for the system of $2+1+1$ dynamical quarks with physical masses \cite{Borsanyi:2021sxv}.

Besides those parameters, $\beta=0.834954$, $g_c=0.3$, $\lambda_4=16$. The critical value of baryon chemical potential is $\mu_{P0}=977 \mathrm{MeV}$. The meaning of this parameter will be interpreted in Subsec. \ref{subsec_NM_EoS}.

\subsubsection{Model \texorpdfstring{\RNum{2}}{2}}
\label{subsec_model_2}

In model \RNum{2}, the potential for the dilation field $\phi(z)$ is
\begin{align}
   & V_{\phi}(\phi)=-12-\left(\frac{3}{2}-\frac{2}{3} c_1\right) {\phi}^2
  \nonumber                                                                                         \\
   & \qquad \qquad -4 c_1 (1+{\phi}^2)^{\frac{c_2}{2}} \left[\sinh(\frac{\phi}{\sqrt{6}})\right]^2,
  \label{V_phi_model_2}
\end{align}
and the function $h_{\phi}(\phi)$ is
\begin{align}
  h_{\phi}(\phi)=\text{sech}(c_3 \phi),
  \label{h_phi_model_2}
\end{align}
where $c_1=1$, $c_2=\frac{1}{2}$, $c_3=0.7$. The function $V_{\phi}(\phi)$ are taken from Ref. \cite{Hamada:2020phg}. The $5$-dimensional Newtonian constant is $G_5=1$. The characteristic energy scale \cite{Critelli:2017oub} of the EMD system is $\Lambda=1 \mathrm{GeV}$.

Besides those parameters, $\beta=1$, $g_c=0$, $\lambda_4=16$. The critical value of baryon chemical potential is $\mu_{P0}=1085 \mathrm{MeV}$.

\subsubsection{Model \texorpdfstring{\RNum{3}}{3}}
\label{subsec_model_3}

In model \RNum{3}, the dilation field $\phi(z)$ is \begin{align}
  \phi(z)=c_1 z^2,
  \label{phi_model_3}
\end{align}
and the function $h_{\phi}$ is
\begin{align}
  h_{\phi}(z)=\mathrm{e}^{c_2 z^2-A_E(z)},
  \label{h_phi_model_3}
\end{align}
where $c_1=1.536 {\mathrm{GeV}}^2$ and $c_2=1.16 {\mathrm{GeV}}^2$. The $5$-dimensional Newtonian constant is $G_5=17.930$. The value of the parameter $c_1$ is taken from Ref. \cite{Zhang:2021itx}, which can produce good results for the glueballs/oddballs spectra compared with those of the lattice QCD simulation, as ``Model \RNum{3}, \RNum{4}(1)'' in Table \RNum{3} in Ref. \cite{Zhang:2021itx}. The form of the function $h_{\phi}(z)$ is inspired by Refs. \cite{He:2013qq,Dudal:2017max}.

Besides those parameters, $\beta=0.7808$, $g_c=0$, $\lambda_4=16$. The critical value of baryon chemical potential is $\mu_{P0}=977 \mathrm{MeV}$.

\subsection{ $T-\mu$ phase diagram for the confinement-deconfinement phase transition }
\label{subsec_T_mu_phase_diagram}

We can get the black\hyp{}hole solution from the EoMs Eq. (\ref{EoMs_2_1})-(\ref{EoMs_2_5}). The temperature of the black\hyp{}hole is
\begin{align}
   & T=\left| \frac{f'(z_h)}{4\pi} \right|,
  \label{temperature}
\end{align}
where $z_h$ is the location of the horizon for the black\hyp{}hole. The entropy density of the black\hyp{}hole is
\begin{align}
   & s_b=\frac{\mathrm{e}^{3 A_E(z_h)}}{\frac{\kappa_5^2}{2\pi} z_h^3}.
  \label{entropy}
\end{align}
According to the AdS/CFT dictionary, the temperature and the entropy density of the QCD matter are $T$ and $s_b$, respectively.

According to the AdS/CFT dictionary, the chemical potential is
\begin{align}
   & \mu=A_t(z=0),
  \label{mu_B}
\end{align}
and the baryon number density is
\begin{align}
   & n_b= \left| \lim_{z\rightarrow 0} {\frac{\partial {{\mathcal{L}}^E}}{\partial \left( \partial_z A_t \right)}} \right|
  \nonumber                                                                                                                                                       \\
   & \quad =-\frac{1}{2 {\kappa_5}^2} \lim_{z\rightarrow 0} \left[\frac{{\mathrm{e}}^{A_E (z)}}{z} h_{\phi}(\phi) \frac{\mathrm{d}}{\mathrm{d} z} A_t (z)\right],
  \label{n_B_index_1}
\end{align}
where ${\mathcal{L}}^E$ is the Lagrangian density in the Einstein frame in Eq. (\ref{Lagrangian_EMD_1}).
From the Euler-Lagrange equation of the field $A_t (z)$:
\begin{align}
  \partial_{M} {\frac{\partial {{\mathcal{L}}^E}}{\partial \left( \partial_M A_t \right)}}-\frac{\partial {{\mathcal{L}}^E}}{\partial A_t}=0,
  \label{Euler_Lagrange_equation_of_At}
\end{align}
we then derive that
\begin{align}
  \partial_z \left[\frac{{\mathrm{e}}^{A_E (z)}}{z} h_{\phi}(\phi) \frac{\mathrm{d}}{\mathrm{d} z} A_t (z)\right]=0.
  \label{euqation_of_conserved_Gauss_charge}
\end{align}
Eq. (\ref{euqation_of_conserved_Gauss_charge}) is actually the EoM of $A_t$ Eq. (\ref{EoMs_2_1}). Thus, we get the conserved Gauss charge associated with the field $A_t$:
\begin{align}
  Q_G=\frac{{\mathrm{e}}^{A_E (z)}}{z} h_{\phi}(\phi) \frac{\mathrm{d}}{\mathrm{d} z} A_t (z).
  \label{conserved_Gauss_charge}
\end{align}
Then we derive
\begin{align}
   & n_b=-\frac{1}{2 {\kappa_5}^2} Q_G.
  \label{n_B_index_2}
\end{align}
Sometimes the following expression of $n_b$ which is related to the quantities at the black\hyp{}hole horizon is more convenient in the calculation:
\begin{align}
   & n_b=-\frac{1}{2 \kappa_5^2}\frac{\mathrm{e}^{A_E(z_h)}}{z_h} h_{\phi}(\phi=\phi(z_h)) {A_t}^{\prime}(z_h).
  \label{n_B_index_3}
\end{align}

The internal energy density and the free energy density are
\begin{align}
   & \epsilon_b=T s_b-P_b+\mu n_b,
  \nonumber                                       \\
   & \mathcal{F}_b=-P_b=\epsilon_b-T s_b-\mu n_b.
  \label{intern_and_free_energy}
\end{align}
The differential relation of the above equations are
\begin{align}
   & \mathrm{d}\epsilon_b=T \mathrm{d}s_b+\mu \mathrm{d}n_b,
  \nonumber                                                                     \\
   & \mathrm{d}\mathcal{F}_b=-\mathrm{d}P_b=-s_b \mathrm{d}T-n_b \mathrm{d}\mu.
  \label{intern_and_free_energy_diff}
\end{align}
From Eq. (\ref{intern_and_free_energy_diff}), we can calculate the pressure $P$ and the internal energy density $\epsilon$. The trace anomaly is
\begin{align}
   & I_b(T,\mu)=\epsilon_b(T,\mu)-3P_b(T,\mu)
  \nonumber                                                              \\
   & \qquad \qquad  \quad = T {s_b}(T,\mu)+\mu {n_b}(T,\mu)-4P_b(T,\mu).
  \label{trace_anom}
\end{align}

The square of the speed of sound is
\begin{align}
  {{c_s}_b}^2=\frac{\mathrm{d}P_b}{\mathrm{d}{\epsilon_b}}.
  \label{square_of_speef_of_sound}
\end{align}

Here all the thermodynamic quantities are labeled with ``b'', which denotes ``background''. The corresponding phase transition is the confinement-deconfinement phase transition.

\begin{figure}[htbp]
  \includegraphics[width=0.9\linewidth,clip=true,keepaspectratio=true]{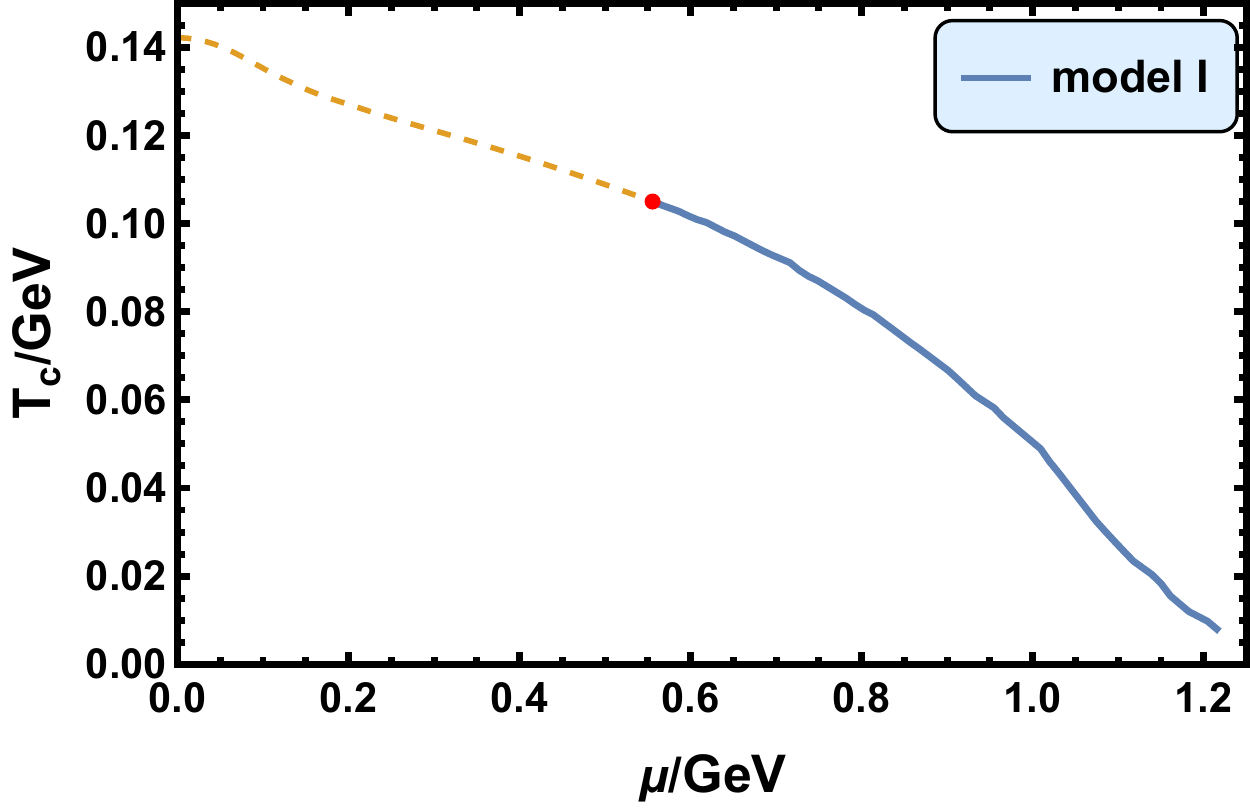}
  \vspace{0.05cm}\\
  \includegraphics[width=0.9\linewidth,clip=true,keepaspectratio=true]{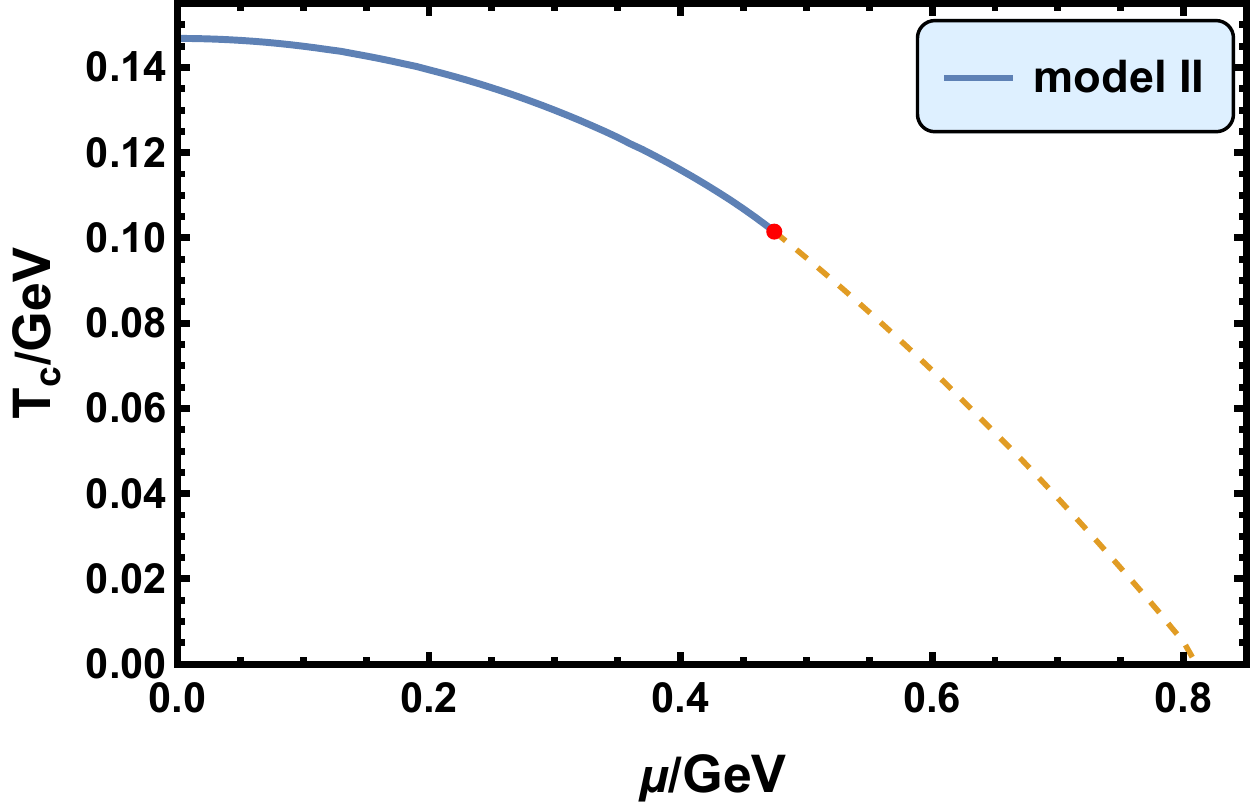}
  \vspace{0.05cm}\\
  \includegraphics[width=0.9\linewidth,clip=true,keepaspectratio=true]{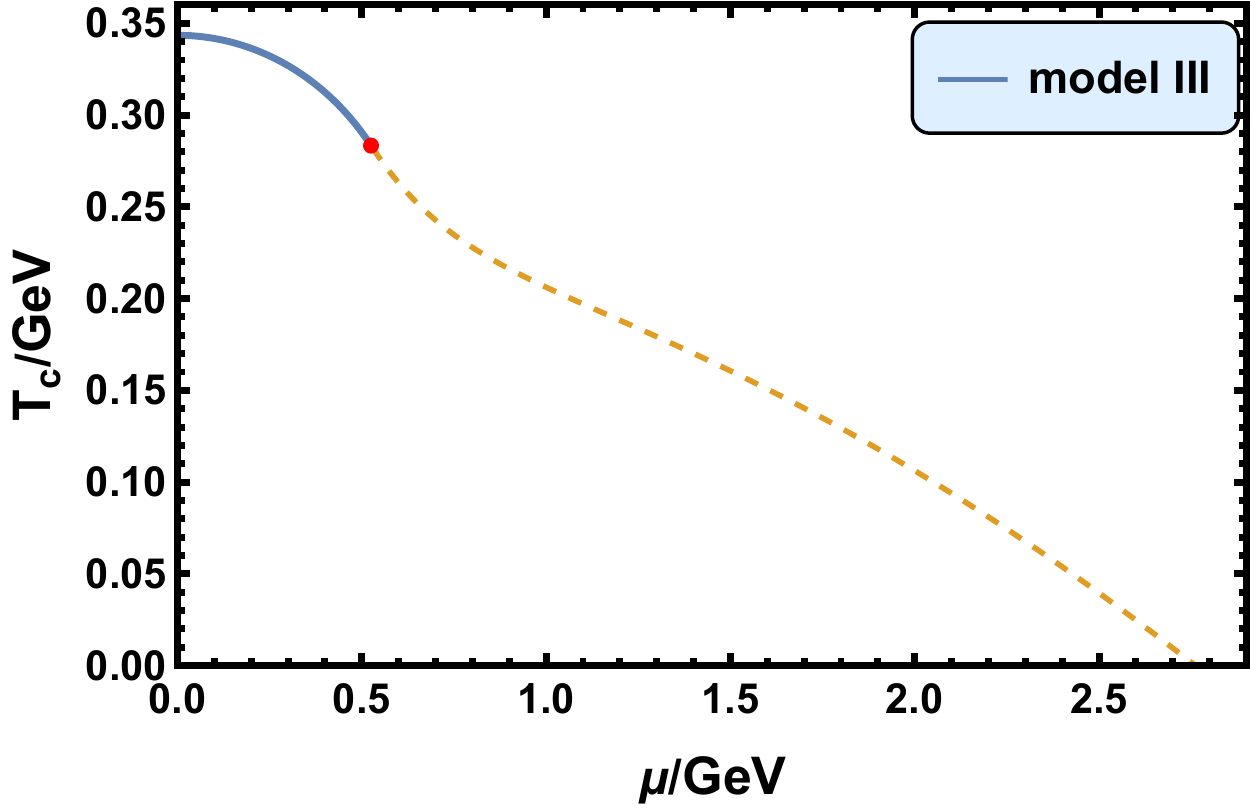}
  \caption{The $T-\mu$ phase diagram of the confinement-deconfinement phase transition in models \RNum{1}, \RNum{2}, and \RNum{3}. The $T_c$ is the critical temperature of the confinement-deconfinement phase transition, $\mu$ is the baryon chemical potential. The blue full line represents that the confinement-deconfinement phase transition is first order transition, and the tan dashed line represents that it's the cross-over behavior. In model \RNum{1} and \RNum{2}, the critical temperature of the cross-over behavior is defined as the temperature where the value of the speed of sound is minimal. In model \RNum{3}, the critical temperature of the cross-over behavior is defined as the temperature where $\lvert \frac{\partial}{\partial T} {z_h}(T,\,\mu) \rvert$ is maximal. The red point is the critical end point (CEP), the position of which is ${(T,\,\mu)}_{\mathrm{CEP}}$. In model \RNum{1}, ${(T,\,\mu)}_{\mathrm{CEP}}=(105.032\mathrm{MeV},\,555.52\mathrm{MeV})$. In model \RNum{2}, ${(T,\,\mu)}_{\mathrm{CEP}}=(101.478\mathrm{MeV},\,474.5\mathrm{MeV})$. In model \RNum{3}, ${(T,\,\mu)}_{\mathrm{CEP}}=(283.359\mathrm{MeV},\,525\mathrm{MeV})$.}
  \label{Tc_mu_three_models}
\end{figure}

The $T-\mu$ phase diagrams for the confinement-deconfinement phase transition in model \RNum{1}, \RNum{2}, and \RNum{3} are shown in in Fig. \ref{Tc_mu_three_models}. $T_c$ is the critical temperature of the confinement-deconfinement phase transition, $\mu$ is the baryon chemical potential. The blue full line represents that the confinement-deconfinement phase transition is first order transition, and the tan dashed line represents that it's the cross-over behavior.

It is noticed that the confinement-deconfinement phase transition is the cross-over behavior at low baryon chemical potential range and a first order phase transition at high baryon chemical potential range in model \RNum{1}. The CEP is located at ${(T,\,\mu)}_{\mathrm{CEP}}=(105.032\mathrm{MeV},\,555.52\mathrm{MeV})$, which is close to the FRG result in Ref. \cite{Fu:2019hdw} and that of the rPNJL model \cite{Li:2018ygx}. The confinement-deconfinement phase transition is a first order phase transition at low baryon chemical potential range and the cross-over behavior at high baryon chemical potential range in model \RNum{2} and model \RNum{3}, which contrasts with that in model \RNum{1}. The CEP is located at ${(T,\,\mu)}_{\mathrm{CEP}}=(101.478\mathrm{MeV},\,474.5\mathrm{MeV})$ in model \RNum{2} and ${(T,\,\mu)}_{\mathrm{CEP}}=(283.359\mathrm{MeV},\,525\mathrm{MeV})$ in model \RNum{3}.

When $\mu=0$, the critical temperature is $142.227\mathrm{MeV}$, $146.803\mathrm{MeV}$, and $343.466\mathrm{MeV}$ in model \RNum{1}, \RNum{2} and \RNum{3}, respectively. When $T=0$, the critical baryon chemical potential is about $1.22\mathrm{GeV}$, $0.81\mathrm{GeV}$, and $2.765\mathrm{GeV}$ in model \RNum{1}, \RNum{2} and \RNum{3}, respectively. In Subsec. \ref{subsec_NM_EoS} we will see that these critical chemical potentials will be changed when we consider the contribution made by the matter part action.

\subsection{Metric deformation in the three models}
\label{subsec_metric_deformat}

To understand the system described by model \RNum{1}, model \RNum{2}, and model \RNum{3}, before we calculate the EoS of the quark matter, we compare the backgrounds in the three models with the ${\mathrm{AdS}}_5$ space\hyp{}time.

\begin{figure}[htbp]
  \includegraphics[width=0.9\linewidth,clip=true,keepaspectratio=true]{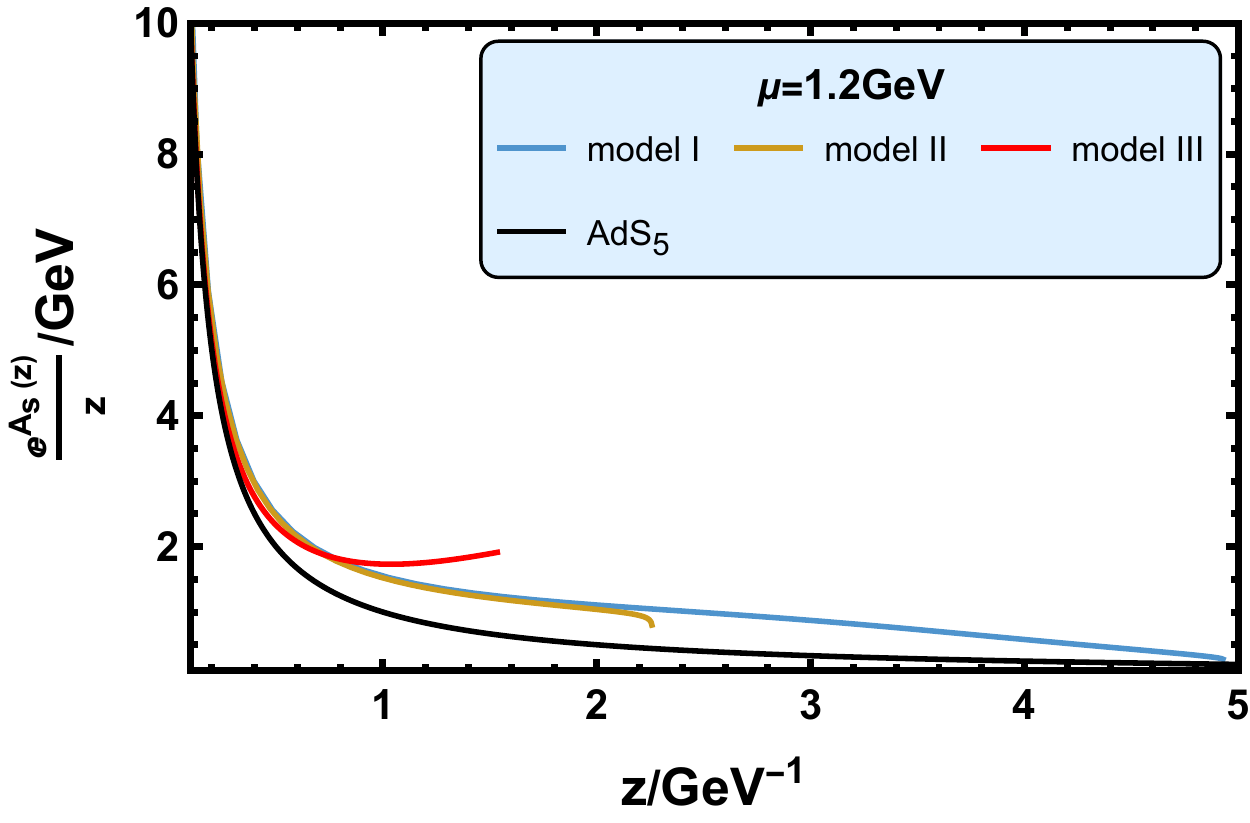}
  \vspace{0.05cm}\\
  \includegraphics[width=0.9\linewidth,clip=true,keepaspectratio=true]{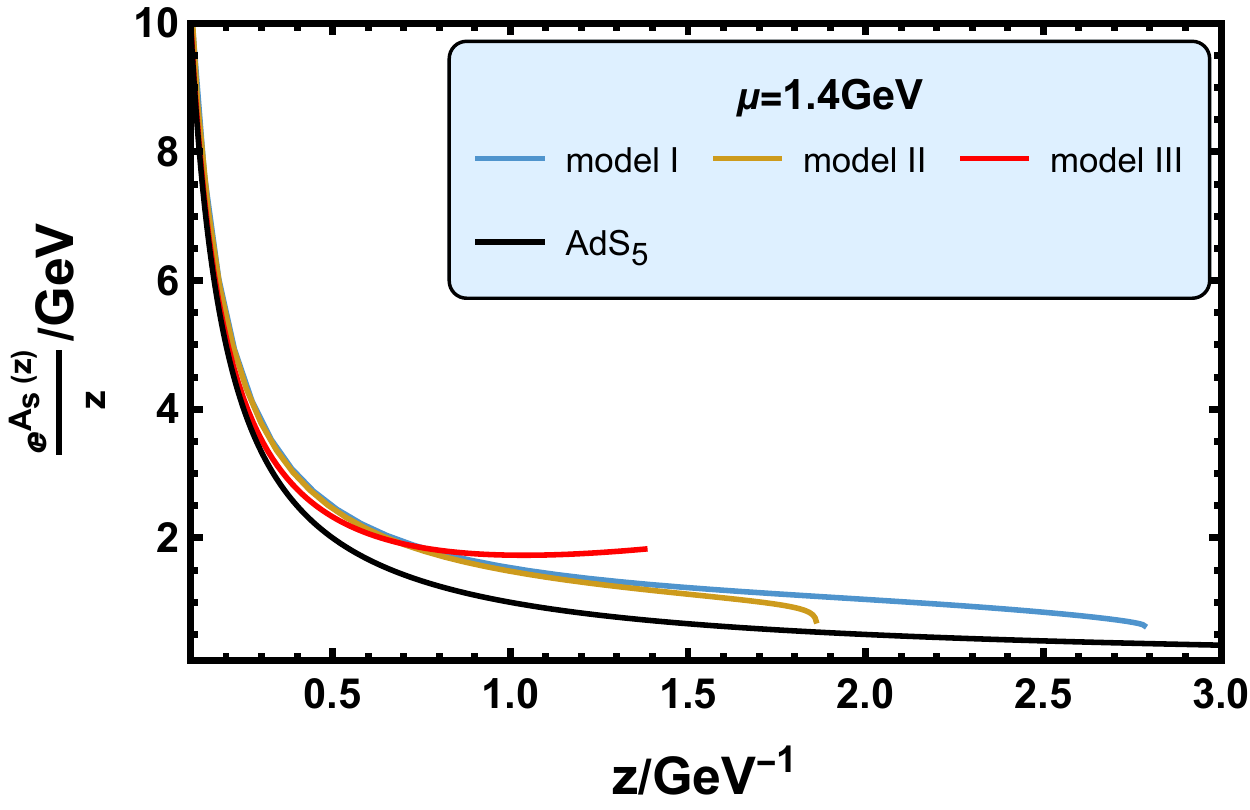}
  \vspace{0.05cm}\\
  \includegraphics[width=0.9\linewidth,clip=true,keepaspectratio=true]{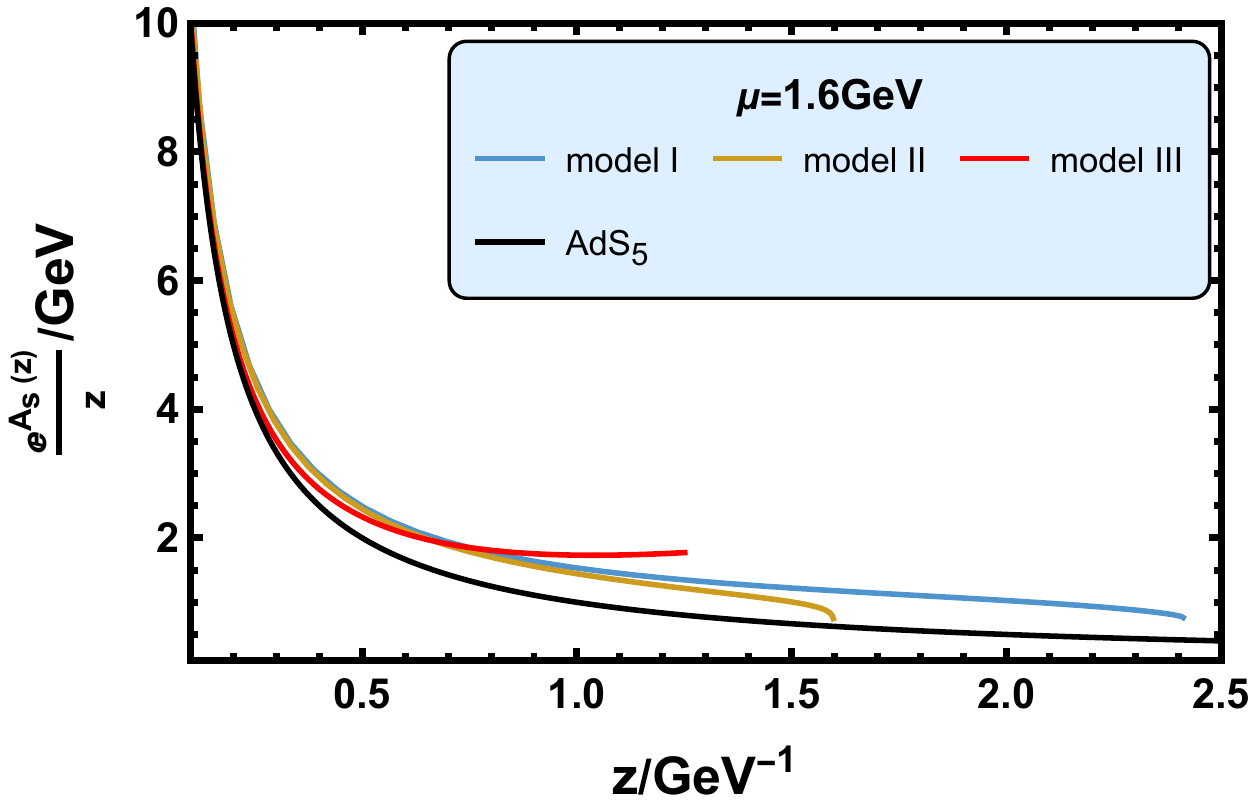}
  \vspace{0.05cm}\\
  \includegraphics[width=0.9\linewidth,clip=true,keepaspectratio=true]{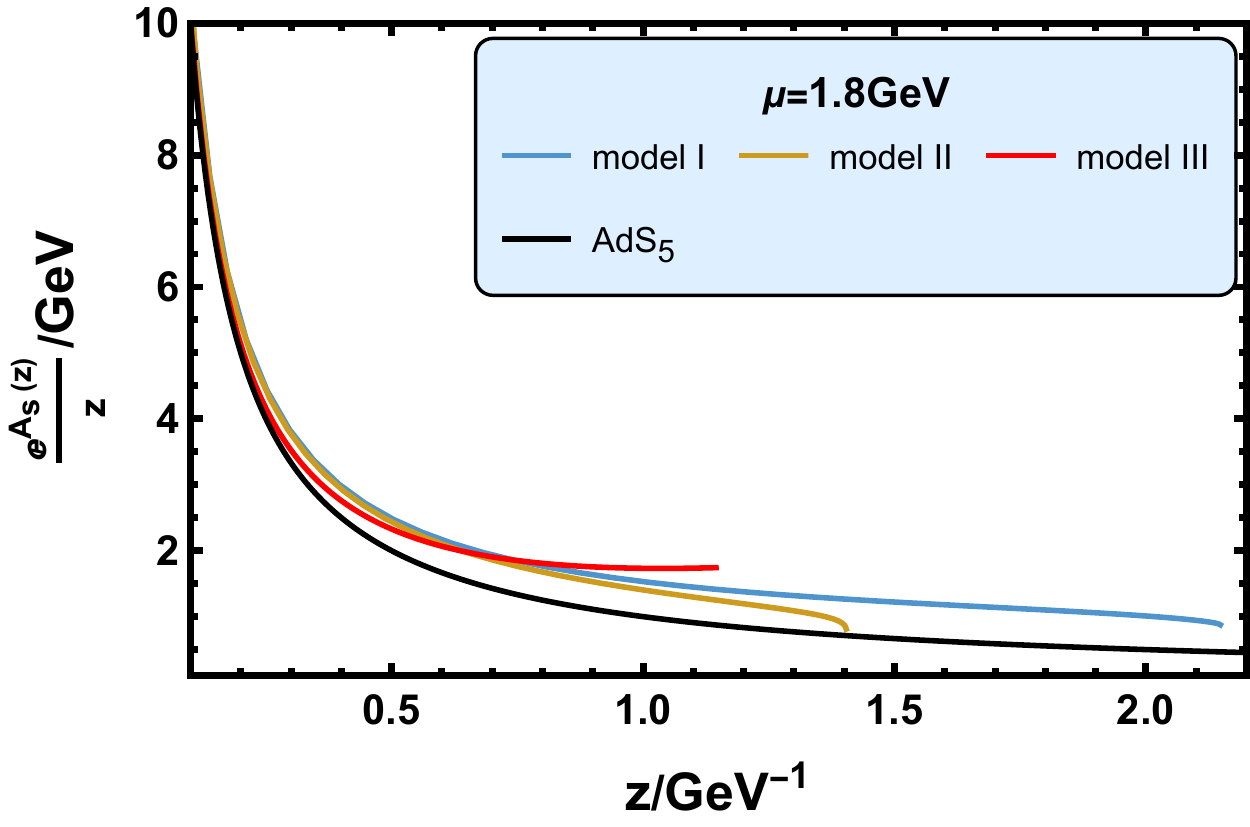}
  \caption{The function $\frac{{\mathrm{e}}^{A_s (z)}}{z}$ in the deformed metric in three models at $\mu=1.2 \mathrm{GeV}$, $\mu=1.4 \mathrm{GeV}$, $\mu=1.6 \mathrm{GeV}$, and $\mu=1.8 \mathrm{GeV}$ at $T=0$.  The steel blue line, the goldenrod line, and the red line represent the function $\frac{{\mathrm{e}}^{A_s (z)}}{z}$ in model \RNum{1}, model \RNum{2}, and model \RNum{3}, respectively. The black line is that in the ${\mathrm{AdS}}_5$ space\hyp{}time, which is $\frac{1}{z}$.}
  \label{deform_metric_func_2}
\end{figure}

In Fig. \ref{deform_metric_func_2}, we plot the function $\frac{{\mathrm{e}}^{A_s (z)}}{z}$ in the deformed metric in three models at four different values of baryon chemical potential: $\mu=1.2 \mathrm{GeV}$, $\mu=1.4 \mathrm{GeV}$, $\mu=1.6 \mathrm{GeV}$, and $\mu=1.8 \mathrm{GeV}$ at $T=0$. In the $AdS_5$ space\hyp{}time, $\frac{{\mathrm{e}}^{A_s (z)}}{z}=\frac{1}{z}$, which is shown by the black solid line. So the deviation of the lines in Fig. \ref{deform_metric_func_2} from the black lines shows the deformation of the metric in the three holographic models.

We can see from Fig. \ref{deform_metric_func_2} that the function $\frac{{\mathrm{e}}^{A_s (z)}}{z}$ in the three models are all located above the $AdS_5$ case. As mentioned in Subsec. \ref{subsec_three_models}, under the parameter values in model \RNum{1}, the EMD system can derive the thermodynamics that is consistent with the lattice QCD result of the system of $2+1+1$ dynamical
quarks with physical masses \cite{Cai:2022omk,Borsanyi:2021sxv}. Thus, we think the background of model \RNum{1} describes the $2+1+1$ flavor system. The parameters in model \RNum{3} can produce good results of the glueballs/oddballs spectra compared with those of the lattice QCD simulation \cite{Zhang:2021itx}. Thus, the background of model \RNum{3} is considered to describe the pure gluon system. Fig. \ref{deform_metric_func_2} shows that the pure gluon system is most deformed in the infrared (IR) range.

In Subsec. \ref{subsec_M_R_relat} we will see that the astrophysical observations of neutron star prefer model \RNum{1} and model \RNum{3} than model \RNum{2}.
So we surmise that the background, the function $\frac{{\mathrm{e}}^{A_s (z)}}{z}$ of which is located between the steel blue line and the red line in Fig. \ref{deform_metric_func_2}, can produce neutron star properties that are consistent with the astrophysical observations.



\section{Equation of state for the cold QCD matter}
\label{sec_EoS}


\subsection{Nuclear matter}
\label{subsec_NM_EoS}
In this subsection, we introduce how to describe nuclear matter in the holographic model.
The total pressure consists of the contribution made by the background part and that made by the matter part:
\begin{align}
   & P_{\mathrm{tot}}=P_{b}+P_{m}.
  \label{P_for_NM}
\end{align}
As discussed in Subsec. \ref{subsec_T_mu_phase_diagram}, the background part of the pressure $P_{b}$ can be calculated from the confined solutions of the EoMs Eq. (\ref{EoMs_2_1})-(\ref{EoMs_2_5}). But the matter part of the pressure $P_{\text{m}}$ is obtained through calculating the on-shell Euclidean action ${S_{m}}_{\text{Euclidean}}^{\text{on-shell}}$:
\begin{align}
   & P_{m}=-{\mathcal{F}}_{m}=-{S_{m}}_{\text{Euclidean}}^{\text{on-shell}}.
  \label{P_for_NM_matter_part}
\end{align}
When we calculate the on-shell Euclidean action for the matter part, we neglect the contribution from the $S_{\text{mesons}}$ and $S_{\text{baryons}}$:
\begin{align}
   & {S_{m}}_{\text{Euclidean}}^{\text{on-shell}}
  \simeq {S_{\chi}}_{\text{Euclidean}}^{\text{on-shell}}.
  \label{Euclidean_on_shell_action_index_1}
\end{align}
As mentioned in Subsec. \ref{subsec_matter_part}, the pressure density from $5$-dimensional fields that dual to the mesons and baryons tower are perturbations compared with that from the field $\chi$. So we think the contribution to the Euclidean on-shell action from $S_{\text{mesons}}$ and $S_{\text{baryons}}$ are negligible.

Now we use model \RNum{1} as the example to illustrate how to describe nuclear matter and then to calculate the EoS of nuclear matter. From Fig. \ref{Tc_mu_three_models}, we can see that there is a intersection point between the $T_c-\mu$ curve and the horizontal axis, the position of which is denoted as $(T=0,\,\mu_c(T=0) \simeq 1.22 \mathrm{GeV})$. When $T=0$ and $\mu<\mu_c(T=0)$, the background is the thermal-gas solutions, the line element for which in Einstein frame is
\begin{align}
  d s^{2}=\frac{L^2 e^{2 A_{E}(z)}}{z^{2}}\left(-d t^{2}+d z^{2}+d y_{1}^{2}+d y_{3}^{2}+d y_{3}^{2}\right).
  \label{metric_Einstein_thermal_gas}
\end{align}
In this case, the values of $s_b$, $n_b$, $P_b$, $\epsilon_b$ are all $0$. Thus, from Eqs. (\ref{P_for_NM})-(\ref{Euclidean_on_shell_action_index_1}) the total pressure is
\begin{align}
  P_{\mathrm{tot}}\simeq -{S_{\chi}}_{\text{Euclidean}}^{\text{on-shell}}.
  \label{P_for_NM_model_1}
\end{align}
When $T=0$ and $\mu>\mu_c(T=0)$, although the thermal-gas solutions also exist, the black\hyp{}hole solutions are more stable, and the corresponding line element in the Einstein frame is Eq. (\ref{metric_Einstein}).

In principle, the calculation should be implemented in the case of $T=0$. However, it's very difficult to calculate the solutions in strict zero temperature. Thus, we calculate the EoS of cold strong interaction matter in a very low temperature, such as $T=0.1085\mathrm{MeV}$ for quark matter. When we calculate the total pressure $P_{\mathrm{tot}}$, we choose a critical value of baryon chemical potential, $\mu_{P0}=977 \mathrm{MeV}$, so that $P_{\mathrm{tot}}(\mu=\mu_{P0})=0$. This value is slightly different from the $939\mathrm{MeV}$ in Sec. ``Baryons from a homogeneous bulk gauge field'' in Ref. \cite{Ishii:2019gta}.

\begin{figure}[htbp]
  \includegraphics[width=\linewidth,clip=true,keepaspectratio=true]{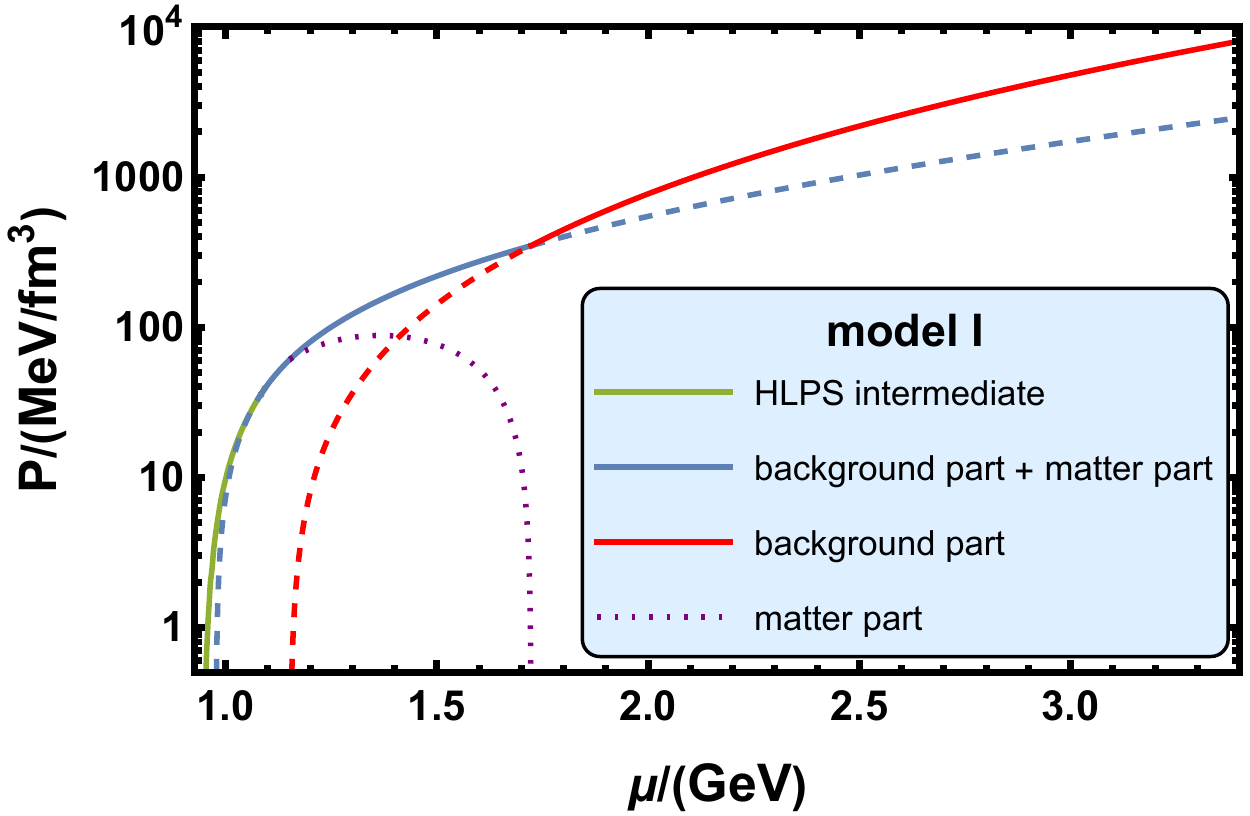}
  \caption{The pressure $P$ of the cold QCD matter as a function of the baryon chemical potential $\mu$. The olive green line, the silver lake blue line, and the red line represent the HLPS intermediate EoS \cite{Hebeler:2013nza}, the holographic result which contains contribution made by the background part and that made by the matter part, and the holographic result which contains only the contribution made by the background part, respectively. The full lines are the thermodynamic stable parts, and the dashed lines are the unstable parts. Additionally, we also plot the dotted purple line which represents the contribution made by the matter part.}
  \label{pressure_to_chemPotentB_model_1}
\end{figure}

We then plot the function $P_{\mathrm{tot}}$ as the silver lake blue line in Fig. \ref{pressure_to_chemPotentB_model_1}. There are also other three different colors in Fig. \ref{pressure_to_chemPotentB_model_1}: the olive green line represents the HLPS intermediate case \cite{Hebeler:2013nza}, the red line, which is calculated in Subsec. \ref{subsec_T_mu_phase_diagram}, represents the holographic result which contains only the contribution made by the background part, and the dotted purple line represents the contribution made by the matter part. There is an intersection between the olive green line and the silver lake blue line, where the baryon chemical potential is $1.084\mathrm{GeV}$. The intersection represents a second-order phase transition, where $P$ and $\frac{\partial P}{\partial_{\mu}} \rvert_{T=0}$ is continuous, but $\frac{\partial^2 P}{\partial_{\mu}^2} \rvert_{T=0}$ are different at both side of this critical values of $\mu$.

Similarly, the intersection between the silver lake blue line and the red line, where $\mu=1.723 \mathrm{GeV}$, represents a first order transition. The value of $\frac{\partial P}{\partial_{\mu}} \rvert_{T=0}$ are different at both side of this critical values of $\mu$. We regard this phase transition as the confinement-deconfinement phase transition in QCD. It is noticed that after considering the contribution made by the matter part action, the critical chemical potential $\mu_c=1.723 \mathrm{GeV}$ is different from that in Subsec. \ref{subsec_T_mu_phase_diagram}. Thus, the silver lake blue line is the EoS for the holographic nuclear matter and the red line is the EoS for the holographic quark matter.

We note that the silver lake blue line is full line only when $1.084\mathrm{GeV} < \mu < 1.723 \mathrm{GeV}$. It means the free energy, in this case, is minimal between the three cases (the olive green line, the silver lake blue line, and the red line) and it is thermodynamic stable when $1.084\mathrm{GeV} < \mu < 1.723 \mathrm{GeV}$. When $\mu < 1.084\mathrm{GeV}$, the nuclear matter with the HLPS intermediate EoS is stable. When $\mu > 1.723 \mathrm{GeV}$, the holographic quark matter is stable.

After we get the pressure of the nuclear matter $P_{NM}=P_{\mathrm{tot}}$, we can calculate the entropy and the baryon number density of the nuclear matter using the differential relation:
\begin{align}
  \mathrm{d}P_{NM}(T,\mu)=s_{NM} \mathrm{d}T+n_{NM} \mathrm{d}\mu.
  \label{pressure_diff_NM}
\end{align}
Then we can derive all other thermodynamic quantities of the nuclear matter. That's why all the thermodynamic quantities here are labeled with ``NM'', which denotes ``nuclear matter''.

\subsection{Hybrid equation of state for neutron stars}
\label{subsec_hybrid_EoS}

\begin{figure}[htbp]
  \includegraphics[width=0.9\linewidth,clip=true,keepaspectratio=true]{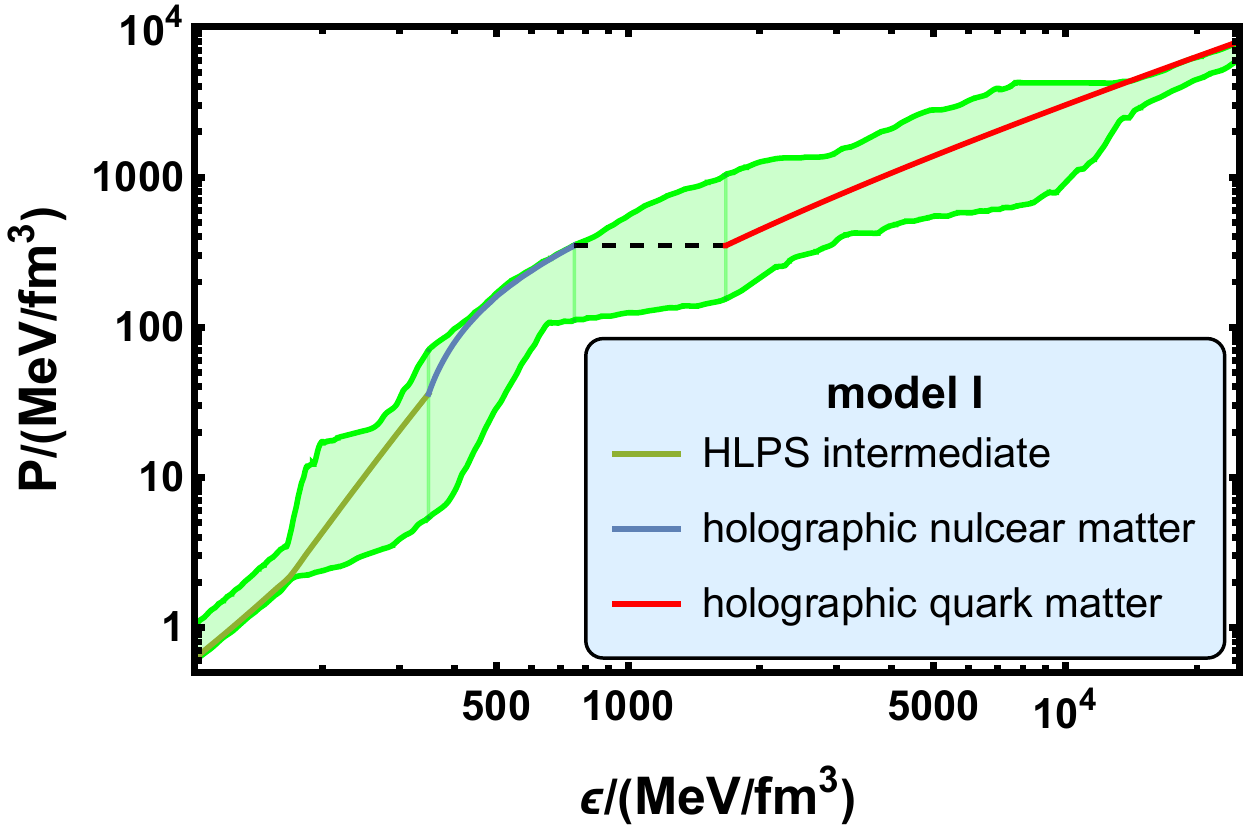}
  \vspace{0cm}\\
  \includegraphics[width=0.9\linewidth,clip=true,keepaspectratio=true]{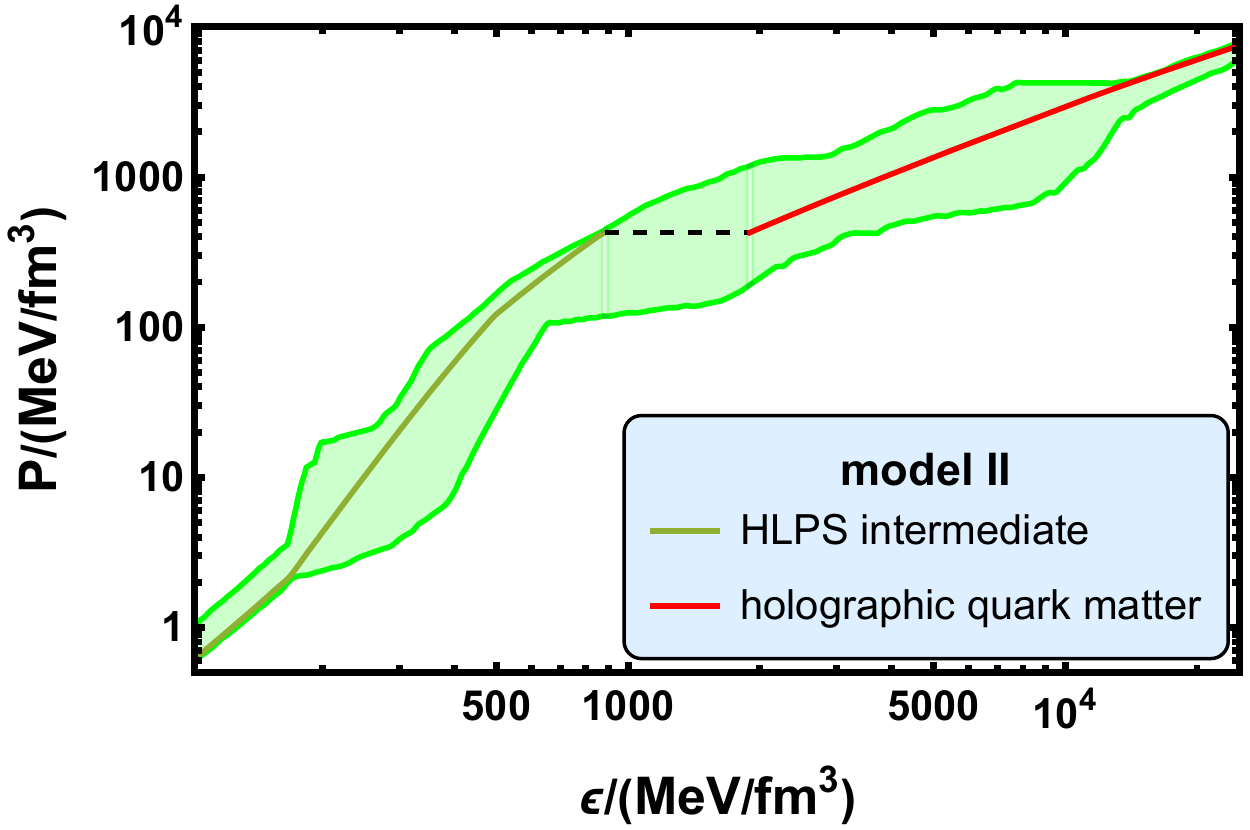}
  \vspace{0cm}\\
  \includegraphics[width=0.9\linewidth,clip=true,keepaspectratio=true]{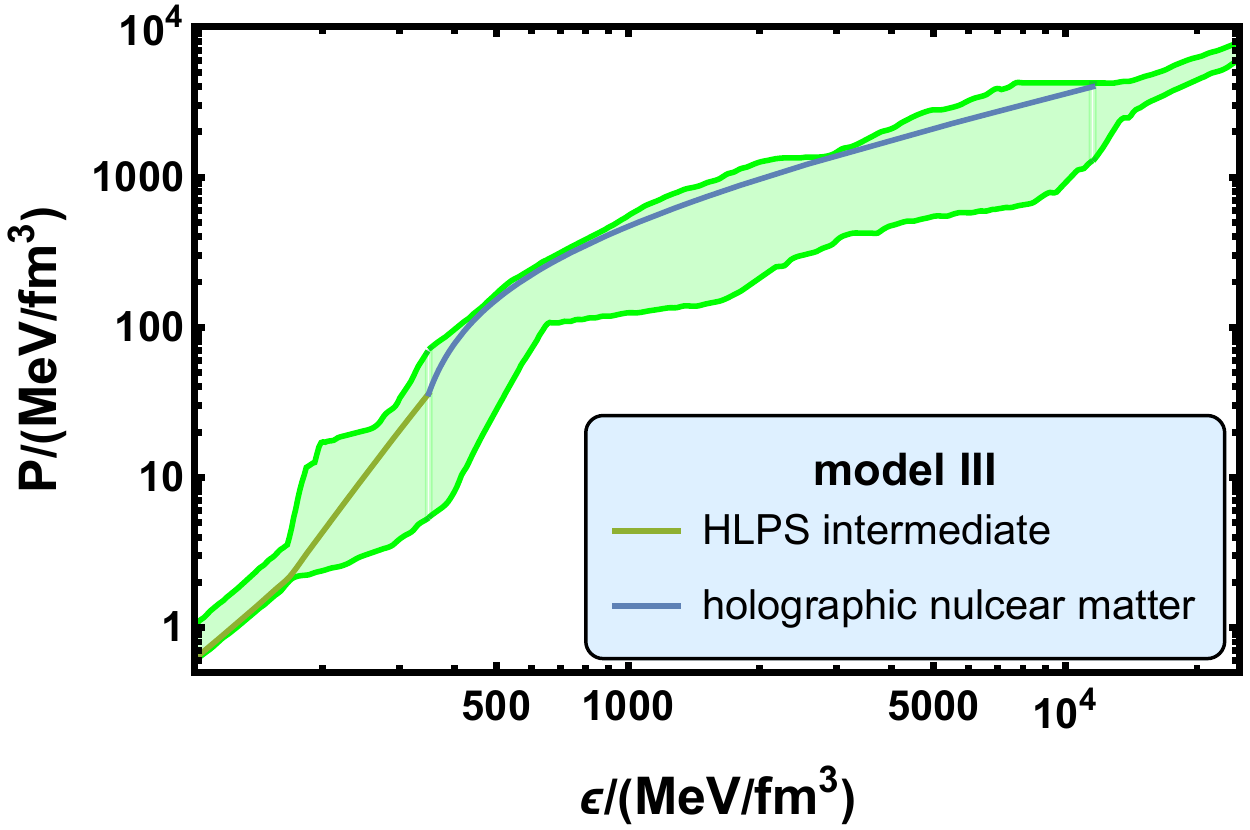}
  \caption{
    The hybrid EoS for the cold QCD matter in the pressure-internal energy density plane calculated in model \RNum{1}, model \RNum{2}, and model \RNum{3}.
    The green band, which is taken from Ref. \cite{Annala:2017llu}, is the ``allowed'' range of the EOS for cold QCD matter. It is defined by the low-density chiral effective theory, the high-density perturbative QCD, and the polytropic interpolations between them, and is constrained by the astrophysics observations.
    In model \RNum{1}, there are three parts in the hybrid EoS. The olive green, the silver lake blue, and the red segment represent the HLPS intermediate EoS \cite{Hebeler:2013nza}, holographic nuclear matter EoS, and holographic quark matter EoS, respectively.
    In model \RNum{2}, there are two parts in the hybrid EoS. The olive green and the red segment represent the HLPS intermediate EoS \cite{Hebeler:2013nza} and holographic quark matter EoS, respectively.
    In model \RNum{3}, there are two parts in the hybrid EoS. The olive green and the silver lake blue segment represent the HLPS intermediate EoS \cite{Hebeler:2013nza}, and holographic nuclear matter EoS, respectively.
  }
  \label{hybrid_EoS_models}
\end{figure}

\begin{figure}[htbp]
  \includegraphics[width=\linewidth,clip=true,keepaspectratio=true]{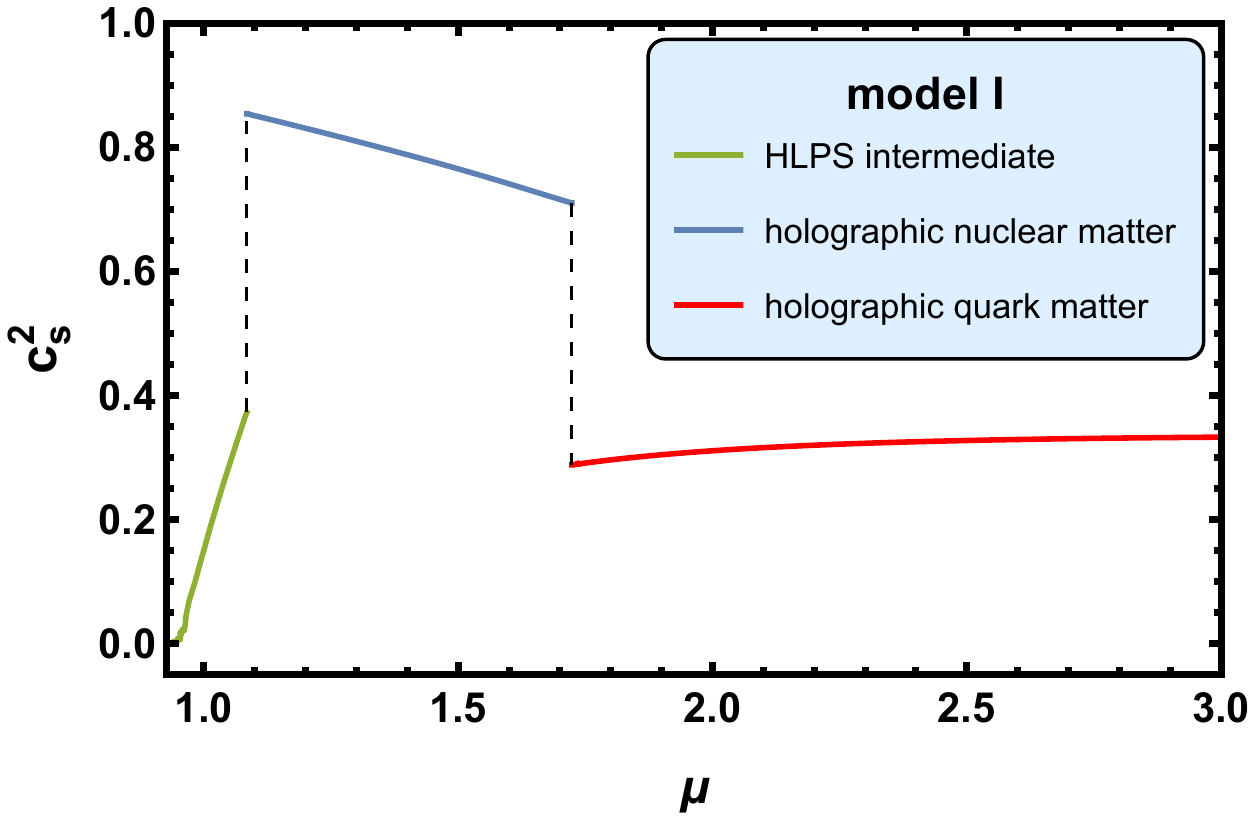}
  \vspace{0.3cm}\\
  \includegraphics[width=\linewidth,clip=true,keepaspectratio=true]{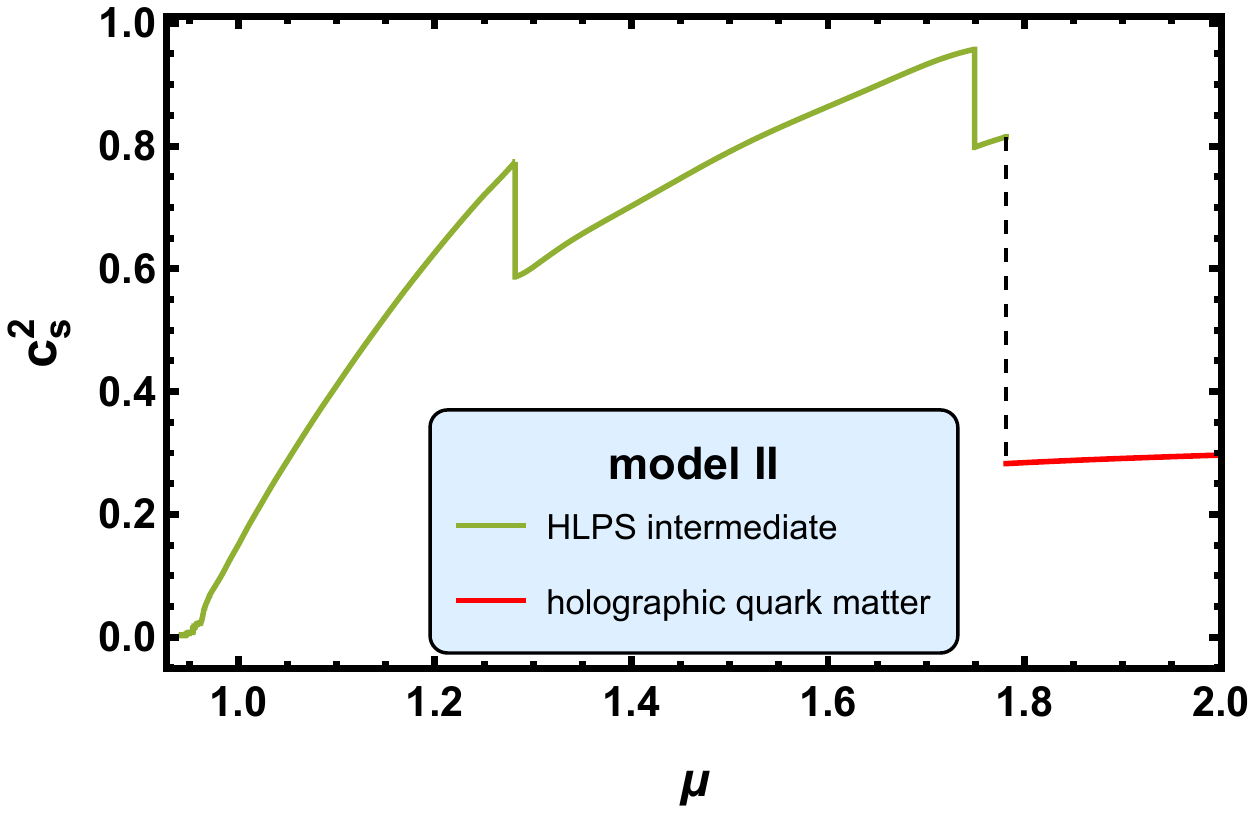}
  \vspace{0.3cm}\\
  \includegraphics[width=\linewidth,clip=true,keepaspectratio=true]{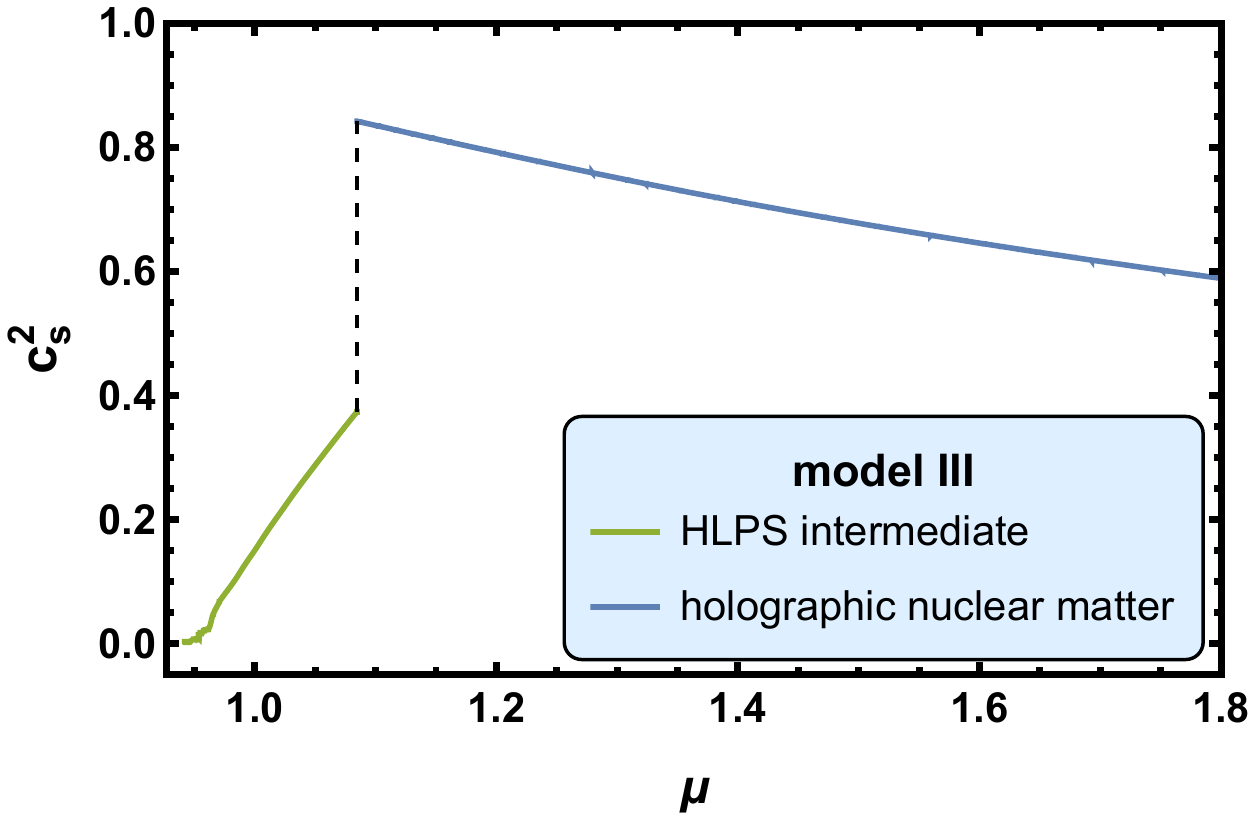}
  \caption{
    The square of the speed of sound $c_s^2$ as a function of the baryon chemical potential $\mu$ for the cold QCD matter calculated in model \RNum{1}, model \RNum{2}, and model \RNum{3}.
    In model \RNum{1}, there are three parts in the panel. The olive green, the silver lake blue, and the red segment correspond to the HLPS intermediate EoS \cite{Hebeler:2013nza}, holographic nuclear matter EoS, and holographic quark matter EoS, respectively.
    In model \RNum{2}, there are two parts in the panel. The olive green and the red segment correspond to the HLPS intermediate EoS \cite{Hebeler:2013nza}, and holographic quark matter EoS, respectively.
    In model \RNum{3}, there are two parts in the panel. The olive green and the silver lake blue segment correspond to the HLPS intermediate EoS \cite{Hebeler:2013nza}, and holographic nuclear matter EoS, respectively.
  }
  \label{speed_of_sound_square_to_chemPotentB_in_three_models}
\end{figure}

\begin{figure}[htbp]
  \includegraphics[width=\linewidth,clip=true,keepaspectratio=true]{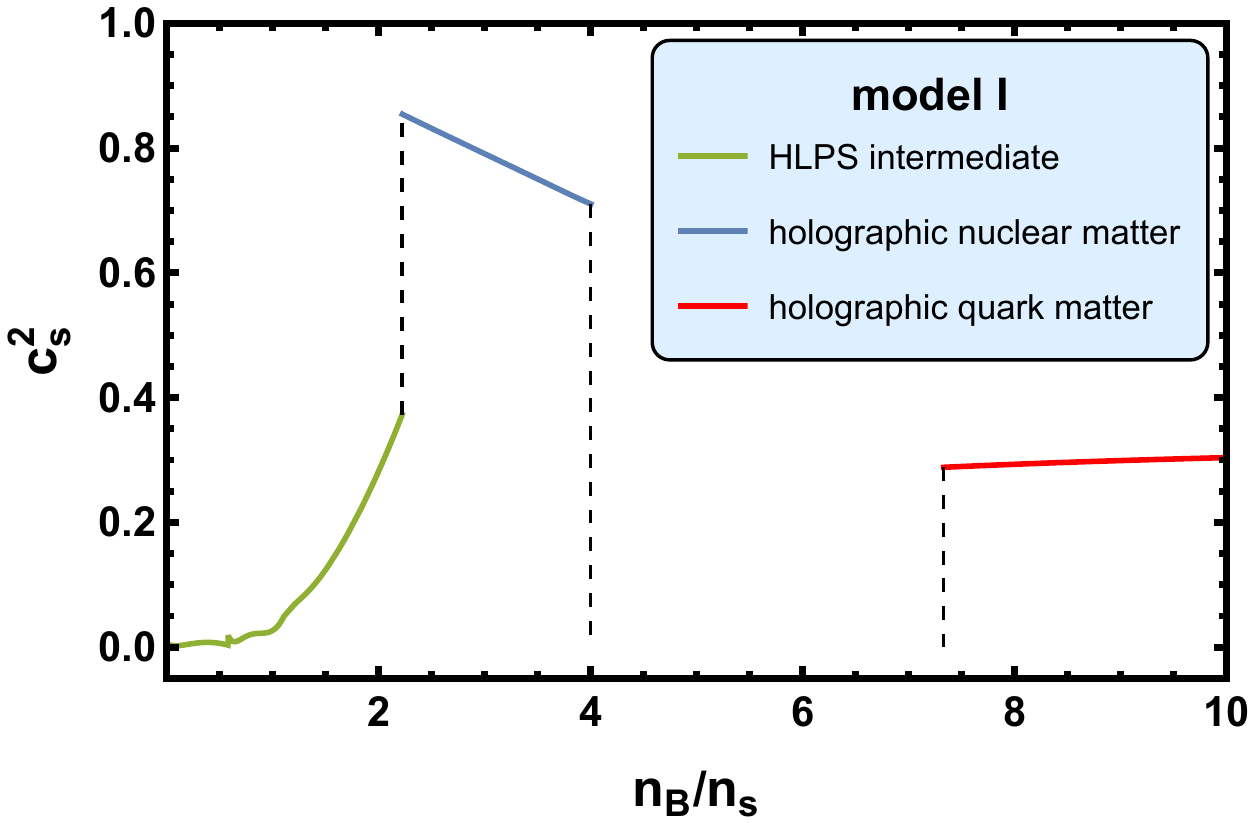}
  \vspace{0.3cm}\\
  \includegraphics[width=\linewidth,clip=true,keepaspectratio=true]{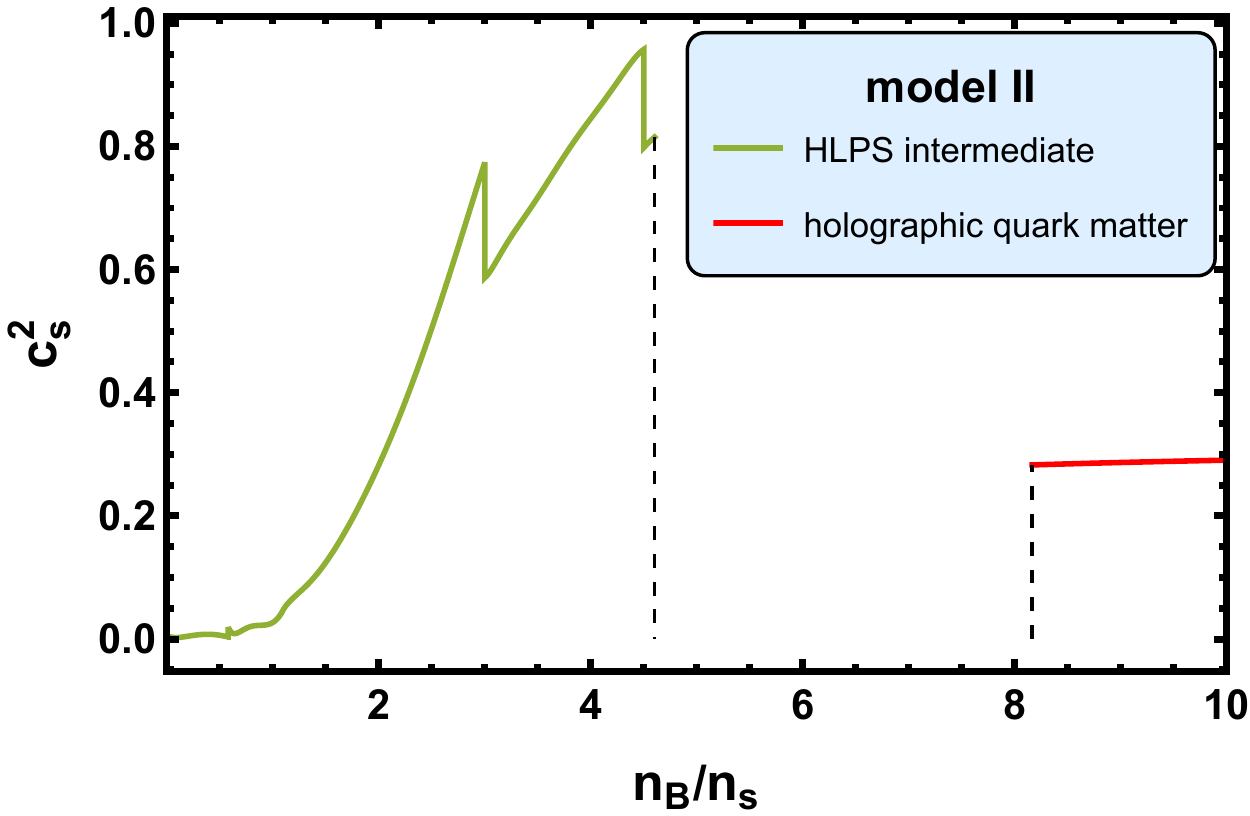}
  \vspace{0.3cm}\\
  \includegraphics[width=\linewidth,clip=true,keepaspectratio=true]{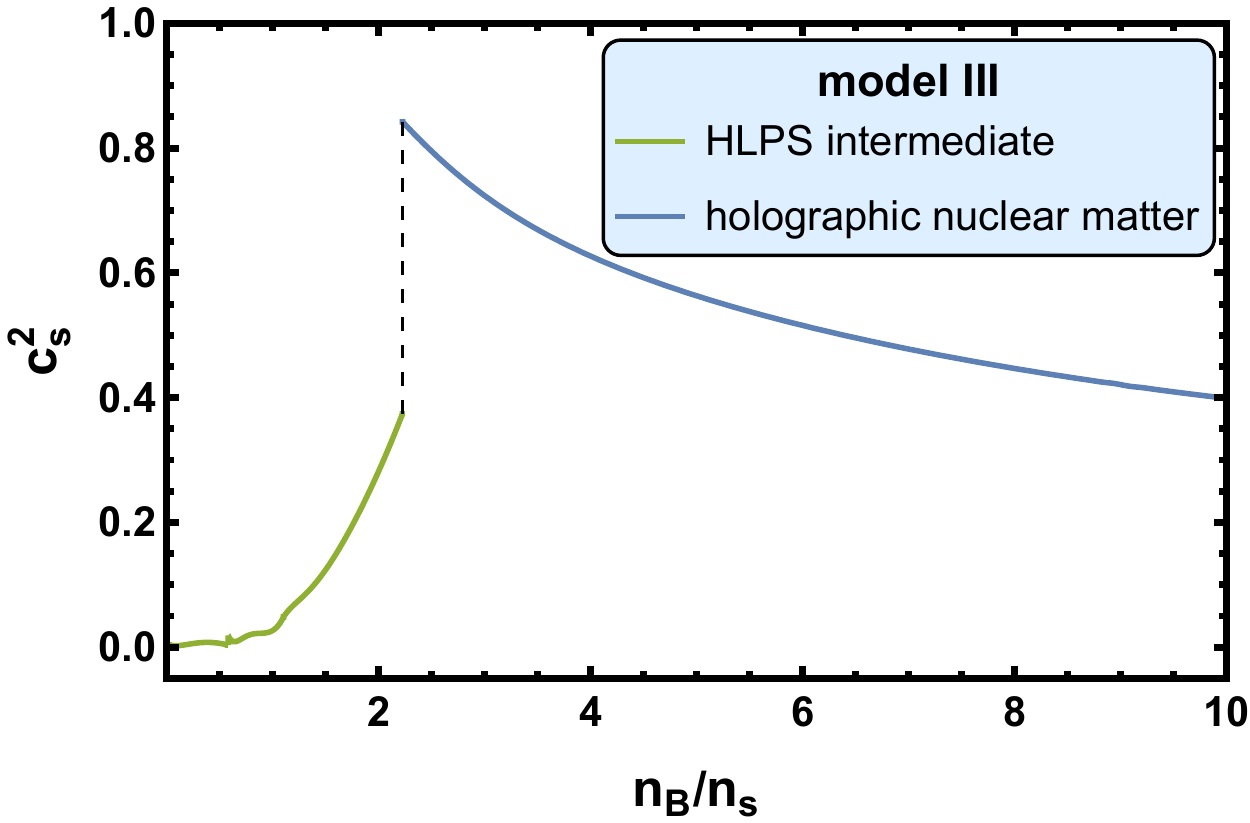}
  \caption{
    The square of the speed of sound $c_s^2$ as a function of the scaled baryon number density $\frac{n_B}{n_s}$ for the cold QCD matter calculated in model \RNum{1}, model \RNum{2}, and model \RNum{3}.
    In model \RNum{1}, there are three parts in the panel. The olive green, the silver lake blue, and the red segment correspond to the HLPS intermediate EoS \cite{Hebeler:2013nza}, holographic nuclear matter EoS, and holographic quark matter EoS, respectively.
    In model \RNum{2}, there are two parts in the panel. The olive green and the red segment correspond to the HLPS intermediate EoS \cite{Hebeler:2013nza}, and holographic quark matter EoS, respectively.
    In model \RNum{3}, there are two parts in the panel. The olive green and the silver lake blue segment correspond to the HLPS intermediate EoS \cite{Hebeler:2013nza}, and holographic nuclear matter EoS, respectively.
  }
  \label{speed_of_sound_square_to_baryDens_in_three_models}
\end{figure}

We show the final results of the hybrid EoS for cold QCD matter from model \RNum{1}, \RNum{2} and \RNum{3} in Fig. \ref{hybrid_EoS_models}, and the corresponding square of the speed of sound in Fig. \ref{speed_of_sound_square_to_chemPotentB_in_three_models} and Fig. \ref{speed_of_sound_square_to_baryDens_in_three_models}.

The hybrid EoS of model \RNum{1} includes three parts, i.e., the HLPS intermediate EoS \cite{Hebeler:2013nza}, the holographic nuclear matter EoS, and the holographic quark matter EoS. The hybrid EoS of model \RNum{2} includes the HLPS intermediate EoS and the holographic quark matter EoS. The hybrid EoS of model \RNum{3} includes the HLPS intermediate EoS and the holographic nuclear matter EoS.

It is observed that the EoSs in the three models all fall in the green band, which is taken from Ref. \cite{Annala:2017llu}, is the ``allowed'' range of the EOS for cold QCD matter. It is defined by the low-density chiral effective theory, the high-density perturbative QCD, and the polytropic interpolations between them, and is constrained by the astrophysics observations.

In Fig. \ref{speed_of_sound_square_to_chemPotentB_in_three_models}, the corresponding square of the speed of sound is shown as a function of the chemical potential. We can observe that the square of the speed of sound increases from a small value at zero chemical potential to a relatively large value more than $0.8$, then decreases to a smaller value at the high baryon chemical potential range. The violation of sound velocity bound in neutron stars has been discussed in much literature, e.g. Refs. \cite{Bedaque:2014sqa,Brandes:2022nxa}. The sound velocity in model \RNum{1} and model \RNum{3} reaches maximum in holographic nuclear matter. However, the maximum sound velocity in model \RNum{2} lies in the HLPS intermediate part.

In Fig. \ref{speed_of_sound_square_to_baryDens_in_three_models}, we show the square of the speed of sound as a function of the scaled baryon number density $\frac{n_B}{n_s}$, where $n_s$ is the nuclear saturation baryon number density.

From Figs. \ref{hybrid_EoS_models}-\ref{speed_of_sound_square_to_baryDens_in_three_models}, we clearly observe that in model \RNum{1}, there are two phase transition. The first one is a second order transition. At this phase transition, the baryon number density $n_B=2.221 n_s$, the baryon chemical potential $\mu=1.084 \mathrm{GeV}$, the internal energy density $\epsilon=349 \mathrm{MeV}/{{\mathrm{fm}}^3}$, and the pressure $P=35.962 \mathrm{MeV}/{{\mathrm{fm}}^3}$. This phase transition connects the HLPS intermediate nuclear matter with the holographic nuclear matter.

The HLPS intermediate nuclear matter and the holographic nuclear matter may be in the same QCD deconfinement phase. So they should be joined together smoothly. However, in practice calculation, the two EoSs come from different models and we have tried our best to use a second order phase transition to join them together. The square of the speed of sound, $c_s^2$, jumps from $0.373$ to $0.854$ through this second order transition. Another possible interpretation for the holographic nuclear matter is that it's in the quarkyonic phase \cite{McLerran:2018hbz,Chen:2019rez} which is in confinement with chiral symmetry restoration. We don't investigate the quarkyonic phase further here and leave it in future work.

The second one is a first order transition, which is interpreted as the confinement-deconfinement phase transition in Subsec. \ref{subsec_NM_EoS} and is located at $\mu=1.723 \mathrm{GeV}$ and $P=349.276 \mathrm{MeV}/{{\mathrm{fm}}^3}$. The value of $n_B$ jumps from $3.999 n_s$ to $7.328 n_s$, the value of $c_s^2$ jumps from $0.710$ to $0.288$, and the value of $\epsilon$ jumps from $753.260 \mathrm{MeV}/{{\mathrm{fm}}^3}$ to $1671.826 \mathrm{MeV}/{{\mathrm{fm}}^3}$ through this first order transition. At $\mu=1.890 \mathrm{GeV}$, the baryon number density $n_B$ reaches $10 n_s$, which is a very high value, and $c_s^2=0.304$. We will see in Subsec. \ref{subsec_M_R_relat} that this first order transition leads to the branch that corresponds to the holographic quark matter is unstable in the mass-radius figure of the NS.

In model \RNum{2}, there is a first order phase transition, which is interpreted as the confinement-deconfinement phase transition and located at $\mu=1.781 \mathrm{GeV}$ and $P=427.250 \mathrm{MeV}/{{\mathrm{fm}}^3}$. The value of $n_B$ jumps from $4.602 n_s$ to $8.163 n_s$, the value of $c_s^2$ jumps from $0.815$ to $0.283$, and the value of $\epsilon$ jumps from $884.477 \mathrm{MeV}/{{\mathrm{fm}}^3}$ to $1899.401 \mathrm{MeV}/{{\mathrm{fm}}^3}$ through this first order transition. At $\mu=1.888 \mathrm{GeV}$, the baryon number density $n_B$ reaches $10 n_s$ and $c_s^2=0.290$. Being similar to the case in model \RNum{1}, this first order phase transition here also leads to the unstable branch that corresponds to the holographic quark matter.

In model \RNum{3}, there is a second order phase transition. The parameter values of this phase transition are $n_B=2.223 n_s$, $\mu=1.085 \mathrm{GeV}$, $\epsilon=349.7 \mathrm{MeV}/{{\mathrm{fm}}^3}$, and $P=36.083 \mathrm{MeV}/{{\mathrm{fm}}^3}$. The square of the speed of sound, $c_s^2$, jumps from $0.374$ to $0.842$ through this second order transition. At $\mu=2.645 \mathrm{GeV}$, the baryon number density $n_B$ reaches $10 n_s$ and $c_s^2=0.399$.

\section{The neutron star}
\label{sec_NS}

\subsection{Mass-radius relation}
\label{subsec_M_R_relat}

The Tolman-Oppenheimer-Volkoff (TOV) equation \cite{Tolman:1934za,Tolman:1939jz,Oppenheimer:1939ne} describes the structure of a spherically symmetric and isotropic object in static gravitational equilibrium:
\begin{align}
   & \frac{\mathrm{d}}{\mathrm{d}r}P(r)+\left[P(r)+\epsilon (r)\right]\frac{G\left[m(r)+4\pi r^3 P(r)\right]}{r^2\left[1-2\frac{G m(r)}{r}\right]}=0,
  \label{TOV_euqation}
\end{align}
where $r$ is the radial coordinate, $P(r)$ is the pressure, and $\epsilon(r)$ is the internal energy density of the material at the radius $r$. $G$ is the Newtonian constant. $m(r)$ is the total mass inside the sphere with radius $r$ measured by the gravitational field felt by a distant observer. It satisfies
\begin{align}
   & \frac{\mathrm{d}}{\mathrm{d}r}m(r)-4\pi r^2 \epsilon(r)=0,
  \label{TOV_m_r}                                               \\
   & m(r=0)=0.
  \label{bound_cond_index_1}
\end{align}
The metric inside the object can be obtained from the element:
\begin{align}
   & {\mathrm{d}s_{\mathrm{in}}}^2=\mathrm{e}^{\nu(r)}{\mathrm{d}t}^2-\frac{{\mathrm{d}r}^2}{1-2\frac{G m(r)}{r}}
  \nonumber                                                                                                        \\
   & \qquad \quad -r^2 \left[{\mathrm{d}\theta_1}^2+ \left(\sin{\theta_1}\right)^2 {\mathrm{d}\theta_2}^2 \right],
  \label{metric_TOV_in_index_1}
\end{align}
where $\nu(r)$ obeys the following equation:
\begin{align}
   & \frac{\mathrm{d}}{\mathrm{d}r}\nu(r)=-\left(\frac{2}{P(r)+\epsilon(r)}\right)\frac{\mathrm{d}}{\mathrm{d}r}P(r).
  \label{metric_TOV_in_index_2}
\end{align}
If the sphere is in the vacuum, the metric outside it is the Schwarzschild metric:
\begin{align}
   & {\mathrm{d}s_{\mathrm{out}}}^2=\left[1-2\frac{G M}{r}\right]{\mathrm{d}t}^2-\frac{{\mathrm{d}r}^2}{1-2\frac{G M}{r}}
  \nonumber                                                                                                               \\
   & \qquad \quad -r^2 \left[{\mathrm{d}\theta_1}^2+ \left(\sin{\theta_1}\right)^2 {\mathrm{d}\theta_2}^2 \right],
  \label{metric_TOV_out}
\end{align}
where $M=m(r=r_B)$ is the total mass of the object measured by the gravitational field felt by a distant observer, $r_B$ is the radius coordinate of the boundary of the object.

To make sure the metric is continuous at the boundary $r=r_B$, we impose the boundary condition
\begin{align}
  \mathrm{e}^{\nu(r=r_B)}=1-2\frac{G M}{r}.
  \label{bound_cond_index_2}
\end{align}
Two other boundary conditions are the zero pressure condition at the boundary of the object
\begin{align}
  P(r=r_B)=0,
  \label{bound_cond_index_3}
\end{align}
and the given internal energy density at $r=0$
\begin{align}
  \epsilon(r=0)=\epsilon_0.
  \label{bound_cond_index_4}
\end{align}

The NS consists of cold QCD matter. We only consider the spherical symmetric and isotropic neutron star in the vacuum without rotation. The EoS of the cold QCD matter can be written in the form
\begin{align}
  \mathcal{C}\left(P(r),\epsilon(r)\right)=0.
  \label{formal_EoS}
\end{align}

\begin{figure}
  \includegraphics[width=\linewidth,clip=true,keepaspectratio=true]{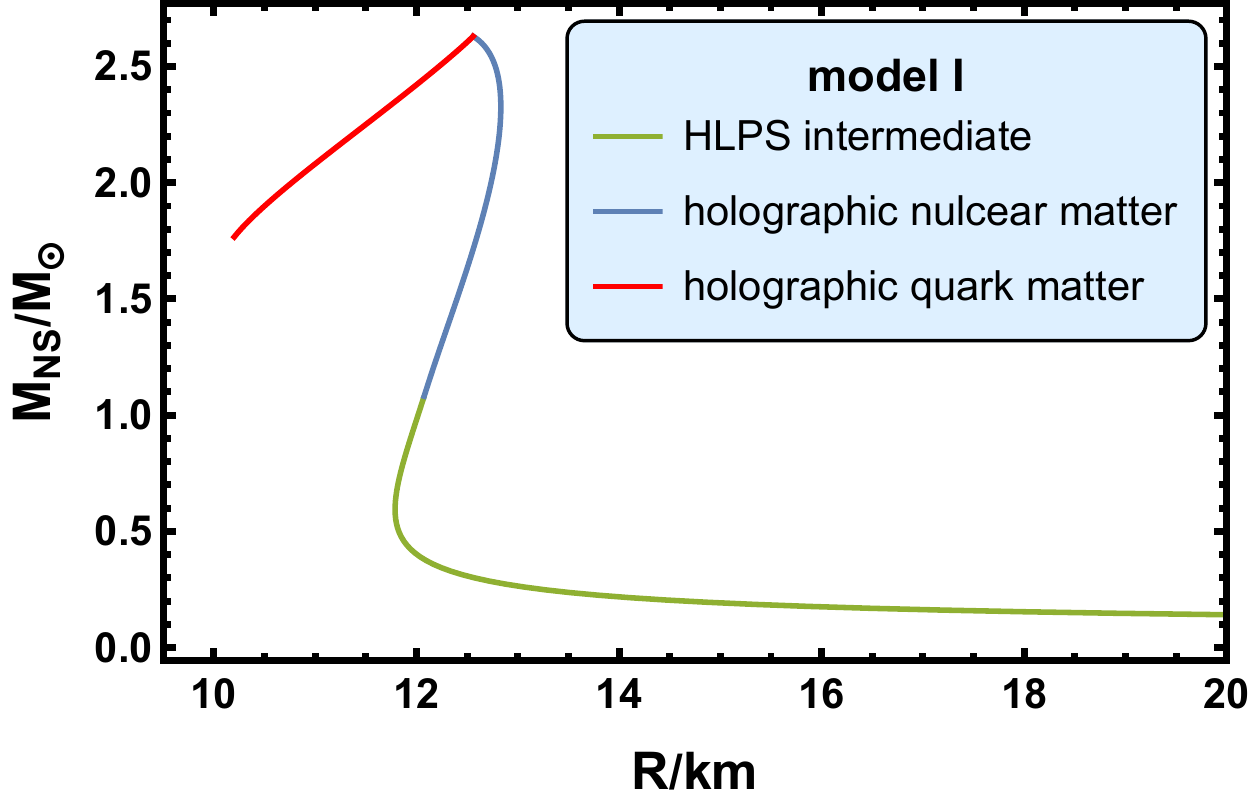}
  \vspace{0.3cm}\\
  \includegraphics[width=\linewidth,clip=true,keepaspectratio=true]{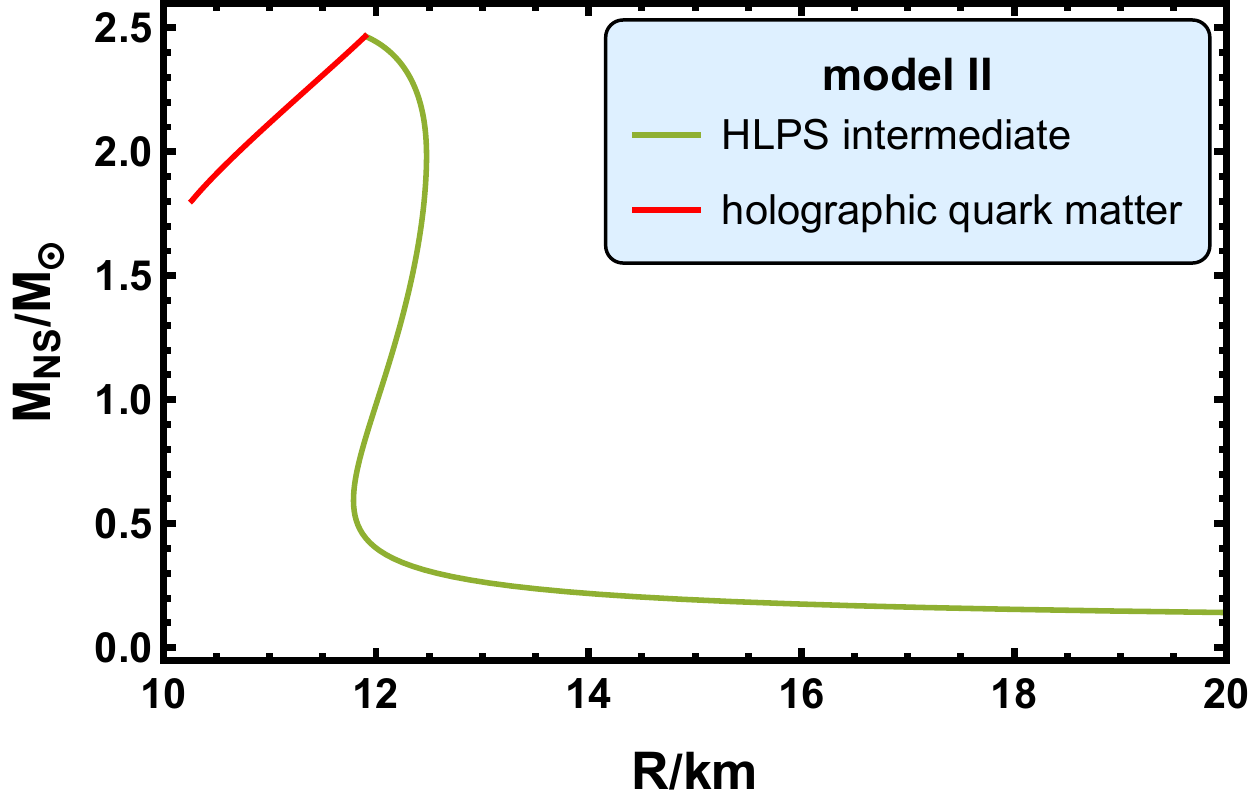}
  \vspace{0.3cm}\\
  \includegraphics[width=\linewidth,clip=true,keepaspectratio=true]{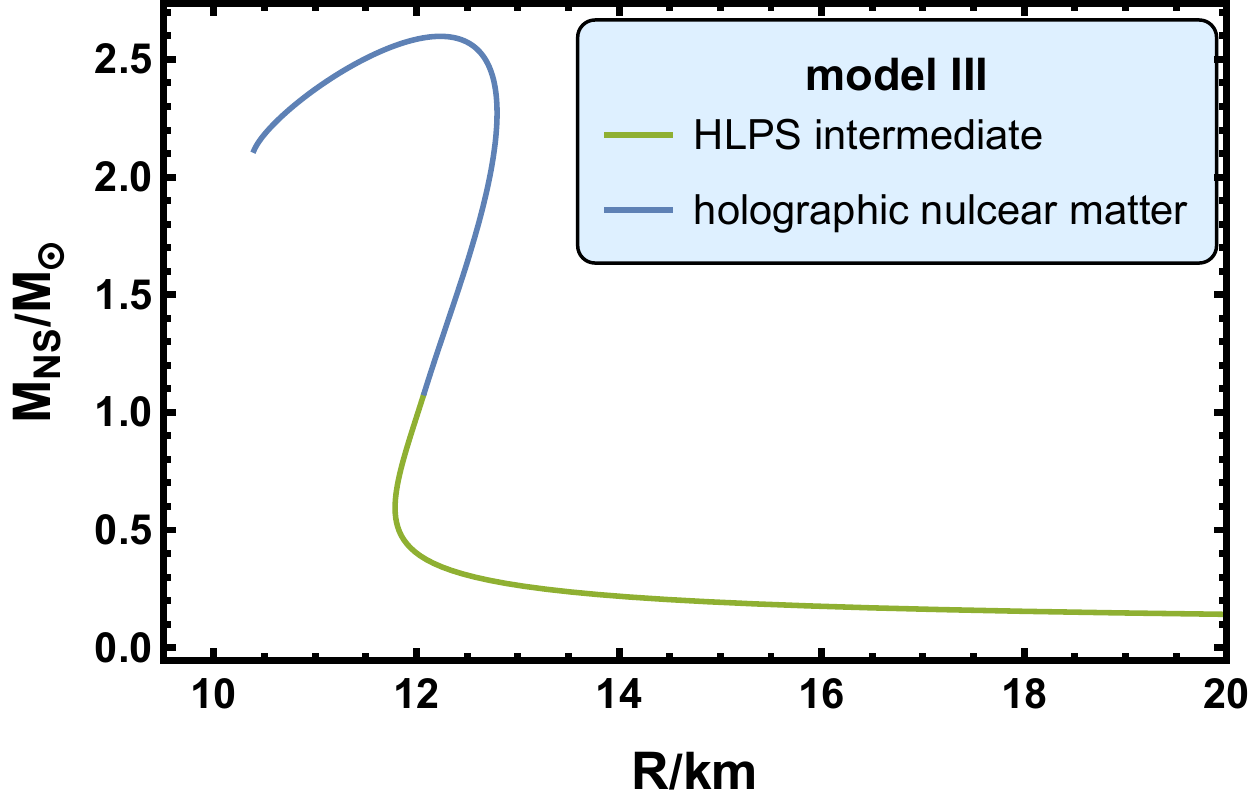}
  \caption{
    The mass-radius relation of the neutron star calculated in model \RNum{1}, model \RNum{2}, and model \RNum{3}.
    In model \RNum{1}, there are three parts in the M-R curve. The olive green, the silver lake blue, and the red segment correspond to the HLPS intermediate EoS \cite{Hebeler:2013nza}, holographic nuclear matter, and holographic quark matter, respectively.
    In model \RNum{2}, there are two parts in the M-R curve. The olive green and the red segment correspond to the HLPS intermediate EoS \cite{Hebeler:2013nza}, and holographic quark matter, respectively.
    In model \RNum{3}, there are two parts in the M-R curve. The olive green and the silver lake blue segment correspond to the HLPS intermediate EoS \cite{Hebeler:2013nza}, and holographic nuclear matter, respectively.
  }
  \label{NS_propert_model_1}
\end{figure}

From Fig \ref{NS_propert_model_1}, we observe that in model \RNum{1}, the radius for the $1.4 M_{\odot}$\hyp{}NS is $12.320 \mathrm{km}$. The max possible radius for the NS is $12.833 \mathrm{km}$, which correspond a $2.32 M_{\odot}$\hyp{}NS. The max possible mass for the NS is $2.630 M_{\odot}$, which correspond a $12.559 \mathrm{km}$\hyp{}NS.

In model \RNum{2}, the radius for the $1.4 M_{\odot}$\hyp{}NS is $12.287 \mathrm{km}$. The max possible radius for the NS is $12.475 \mathrm{km}$, which correspond a $1.98 M_{\odot}$\hyp{}NS. The max possible mass for the NS is $2.466 M_{\odot}$, which correspond a $11.902 \mathrm{km}$\hyp{}NS.

In model \RNum{3}, the radius for the $1.4 M_{\odot}$\hyp{}NS is $12.319 \mathrm{km}$. The max possible radius for the NS is $12.795 \mathrm{km}$, which correspond a $2.27 M_{\odot}$\hyp{}NS. The max possible mass for the NS is $2.598 M_{\odot}$, which correspond a $12.24 \mathrm{km}$\hyp{}NS.

In conclusion, the max possible mass for the neutron star is about $2.5 M_{\odot}$ and the radius is about $12 \mathrm{km}$ then. It is noticed that the branch that corresponds to the holographic quark matter is always unstable, and the holographic nuclear matter can produce the stable branch. These results indicate that even in the core of the NS, the matter is still in the confinement phase and the quark matter is not favored. This is consistent with the conclusion in Ref. \cite{Jokela:2021vwy}, in which the authors combined the input from the gauge\hyp{}gravity duality and the input from various $ab$-$initio$ methods for the nuclear matter at low density and analyzed families of hybrid equations of state of cold QCD matter. Their prediction is that all neutron stars are fully hadronic without quark matter cores.

\subsection{Tidal deformability}
\label{tidal_deform}
At last, we investigate the tidal deformability of the NSs. We consider a static, spherically symmetric object in a static external quadrupolar tidal field ${\mathcal{E}}_{ij}$ \cite{Flanagan:2007ix,Hinderer:2007mb,Hinderer:2009ca}. Then the tidal deformability $\lambda$ is defined to linear order as
\begin{align}
  Q_{ij}=-\lambda {\mathcal{E}}_{ij},
  \label{dimensional_tidal_deform}
\end{align}
where $Q_{ij}$ is the quadrupole moment.
The dimensional tidal deformability is also related to the $l=2$ dimensionless tidal Love number $k_2$:
\begin{align}
  k_2=\frac{3}{2}\lambda G R^{-5}.
  \label{l_2_Love_number_index_1}
\end{align}

The asymptotic expansion of the $tt$ component of the full metric outside the object at a large distance $r$ from the object is
\begin{align}
   & {g_{\mathrm{out}}}_{tt}=1-2\frac{G M}{r}-\frac{3 G Q_{ij}}{r^3}\frac{x^i}{r}\frac{x^j}{r}+\cdots
  \nonumber                                                                                           \\
   & \qquad \quad +{\mathcal{E}}_{ij}r^2\frac{x^i}{r}\frac{x^j}{r}+\cdots.
  \label{asymp_expans_of_g_tt}
\end{align}
The right hand side (r.h.s.) of Eq. (\ref{asymp_expans_of_g_tt}) includes two leading-order terms of the perturbative expansions. The second one describes an external tidal field increasing with $r^2$ and the first one describes the resulting tidal distortion decreasing with $r^{-3}$.

Introducing a linear $l=2$ perturbation on the object leads to a static, even-parity perturbation on the metric \cite{Hinderer:2007mb,Hinderer:2009ca} in the Regge-Wheeler gauge \cite{Regge:1957td}:
\begin{align}
   & {\mathrm{d}s_{\mathrm{perturb}}}^2={\mathrm{e}}^{\nu (r)}\left[1+H(r)Y_{20}(\theta_1,\theta_2)\right]{\mathrm{d}t}^2
  \nonumber                                                                                                                                                     \\
   & \qquad \quad -\frac{\left[1-H(r)Y_{20}(\theta_1,\theta_2)\right]}{1-2\frac{G m(r)}{r}}{\mathrm{d}r}^2
  \nonumber                                                                                                                                                     \\
   & \qquad \quad -r^2 \left[1-K(r)Y_{20}(\theta_1,\theta_2)\right] \left[{\mathrm{d}\theta_1}^2+ \left(\sin{\theta_1}\right)^2 {\mathrm{d}\theta_2}^2 \right],
  \label{perturb_metric_in}
\end{align}
where
\begin{align}
  \frac{\mathrm{d}}{\mathrm{d}r}K(r)=\frac{\mathrm{d}}{\mathrm{d}r}H(r)+H(r)\frac{\mathrm{d}}{\mathrm{d}r} \nu (r).
  \label{K_H_relat}
\end{align}
The corresponding perturbations of the stress-energy tensor of the perfect fluid are
\begin{align}
  \delta T_{0}^{0}=-\delta \epsilon (r) Y_{20}(\theta_1, \theta_2)
  \label{perturb_T_00}
\end{align}
\begin{align}
  \delta T_{i}^{i}=\delta P(r) Y_{20}(\theta_1, \theta_2).
  \label{perturb_T_ii}
\end{align}
The function $H(r)$ obeys the following differential equation:
\begin{align}
   & \Bigg\{
  -\frac{6}{r^2 \left(1-2\frac{G m(r)}{r}\right)}-\frac{1}{2}\left[\frac{\mathrm{d}}{\mathrm{d}r} \nu (r)\right]^2+\frac{{\mathrm{d}}^2}{\mathrm{d}r^2} \nu(r)
  \nonumber                                                                                                                                                                                                                                                              \\
   & +\frac{3}{r}\frac{G \left[ r \frac{\mathrm{d}}{\mathrm{d}r}m(r) - m(r) \right]}{r^2\left[1-2\frac{G m(r)}{r}\right]}+\frac{7}{2r}\frac{\mathrm{d}}{\mathrm{d}r} \nu (r)
  \nonumber                                                                                                                                                                                                                                                              \\
   & -\frac{\mathrm{d}}{\mathrm{d}r} \nu (r) \frac{G \left[ r \frac{\mathrm{d}}{\mathrm{d}r}m(r) - m(r) \right]}{r^2\left[1-2\frac{G m(r)}{r}\right]}
  \nonumber                                                                                                                                                                                                                                                              \\
   & +\frac{1}{2 r}\frac{\frac{\mathrm{d}}{\mathrm{d}r} \epsilon (r)}{\frac{\mathrm{d}}{\mathrm{d}r} P(r)}\bigg\{\frac{\mathrm{d}}{\mathrm{d}r}\nu (r)+\frac{2 G \left[ r \frac{\mathrm{d}}{\mathrm{d}r}m(r) - m(r) \right]}{r^2\left[1-2\frac{G m(r)}{r}\right]}\bigg\}
  \Bigg\}H(r)
  \nonumber                                                                                                                                                                                                                                                              \\
   & +\frac{1}{2}\left\{
  \frac{4}{r}+\frac{\mathrm{d}}{\mathrm{d}r}\nu (r)-\frac{2 G \left[ r \frac{\mathrm{d}}{\mathrm{d}r}m(r) - m(r) \right]}{r^2\left[1-2\frac{G m(r)}{r}\right]}
  \right\}\frac{\mathrm{d}}{\mathrm{d}r}H(r)
  \nonumber                                                                                                                                                                                                                                                              \\
   & +\frac{{\mathrm{d}}^2}{\mathrm{d}r^2}H(r)=0.
  \label{ODE_of_H}
\end{align}

Being similar to the method of solving the moment of inertia in the slow rotation approximation \cite{1969ApJ...158..719H}, we can calculate the tidal perturbation for a given EoS $\mathcal{C}\left(P(r),\epsilon(r)\right)=0$ \cite{Hinderer:2009ca}.
\begin{align}
   & \beta(r)-\frac{\mathrm{d}}{\mathrm{d}r}H(r)=0,
  \label{TOV_index_3}                                                                                                           \\
   & \frac{\mathrm{d}}{\mathrm{d}r}\beta(r)-\frac{2H(r)}{\left[1-2\frac{G m(r)}{r}\right]}\Bigg\{
  -2\pi G \bigg[
    5 \epsilon(r)+9 P(r)
  \nonumber                                                                                                                     \\
   & \qquad +\frac{\frac{\mathrm{d}}{\mathrm{d}r}\epsilon(r)}{\frac{\mathrm{d}}{\mathrm{d}r}P(r)}\left[P(r)+\epsilon (r)\right]
    \bigg]
  +\frac{3}{r^2}
  \nonumber                                                                                                                     \\
   & \qquad+\frac{2}{1-2\frac{G m(r)}{r}}G^2 \left[\frac{M}{r^2}+4\pi r P(r)\right]^2
  \Bigg\}
  \nonumber                                                                                                                     \\
   & \quad-\frac{2\beta(r)}{r\left[1-2\frac{G m(r)}{r}\right]}\Bigg\{-1+\frac{G m(r)}{r}
  \nonumber                                                                                                                     \\
   & \qquad+2\pi G r^2 \left[\epsilon(r)-P(r)\right]
  \Bigg\}=0.
\end{align}
The boundary conditions are
\begin{align}
   & H(r \rightarrow 0)\rightarrow {a_0}r^2,
  \label{bound_cond_index_5}                    \\
   & \beta(r \rightarrow 0)\rightarrow 2 a_0 r.
  \label{bound_cond_index_6}
\end{align}
The constant $a_0$ determines the extent of deformation of the object and will be canceled in the calculation of the Love number. Eq. (\ref{TOV_index_3})-(\ref{bound_cond_index_6}) can be solved together with the TOV system Eq. (\ref{TOV_euqation}), Eq. (\ref{TOV_m_r}), Eq. (\ref{metric_TOV_in_index_2}), Eq. (\ref{formal_EoS}), Eq. (\ref{bound_cond_index_1}), Eq. (\ref{bound_cond_index_2}), Eq. (\ref{bound_cond_index_3}), and Eq. (\ref{bound_cond_index_4}).

\begin{figure}[htbp]
  \includegraphics[width=\linewidth,clip=true,keepaspectratio=true]{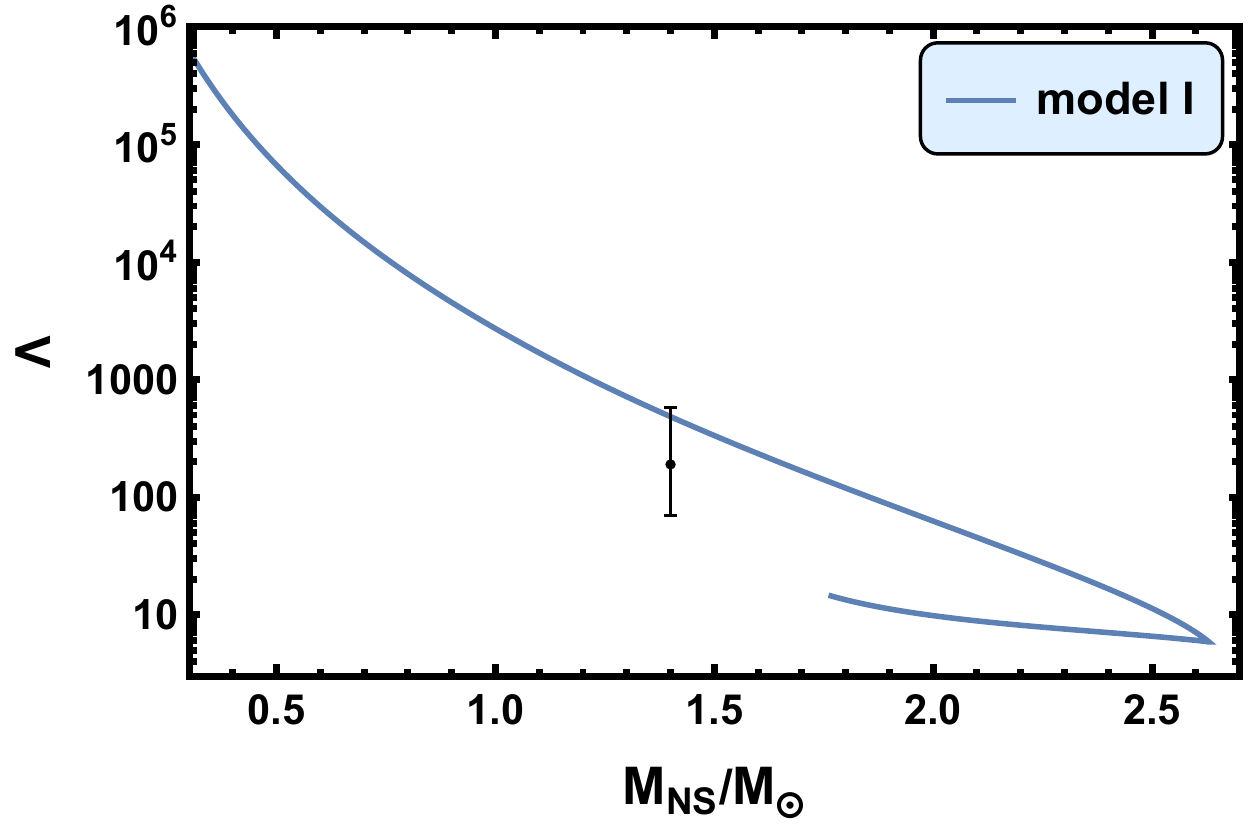}
  \vspace{0.3cm}\\
  \includegraphics[width=\linewidth,clip=true,keepaspectratio=true]{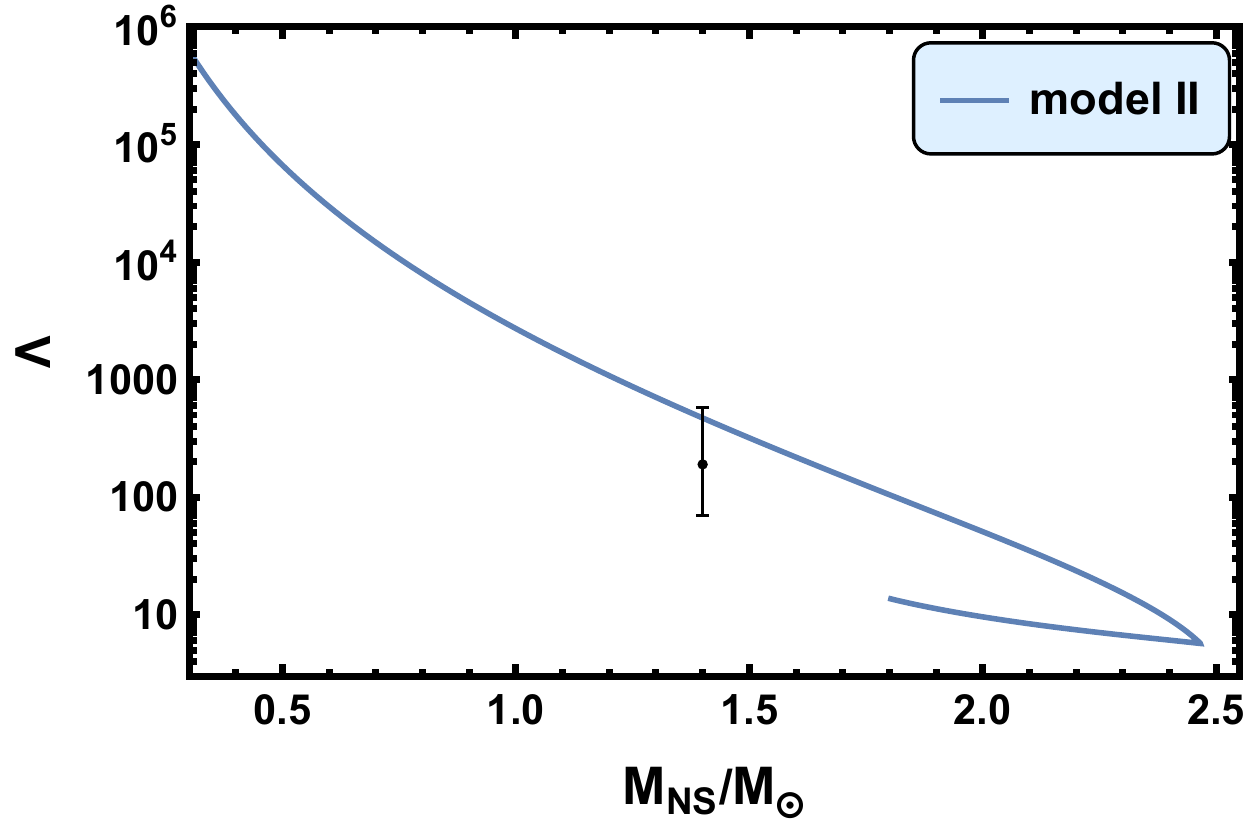}
  \vspace{0.3cm}\\
  \includegraphics[width=\linewidth,clip=true,keepaspectratio=true]{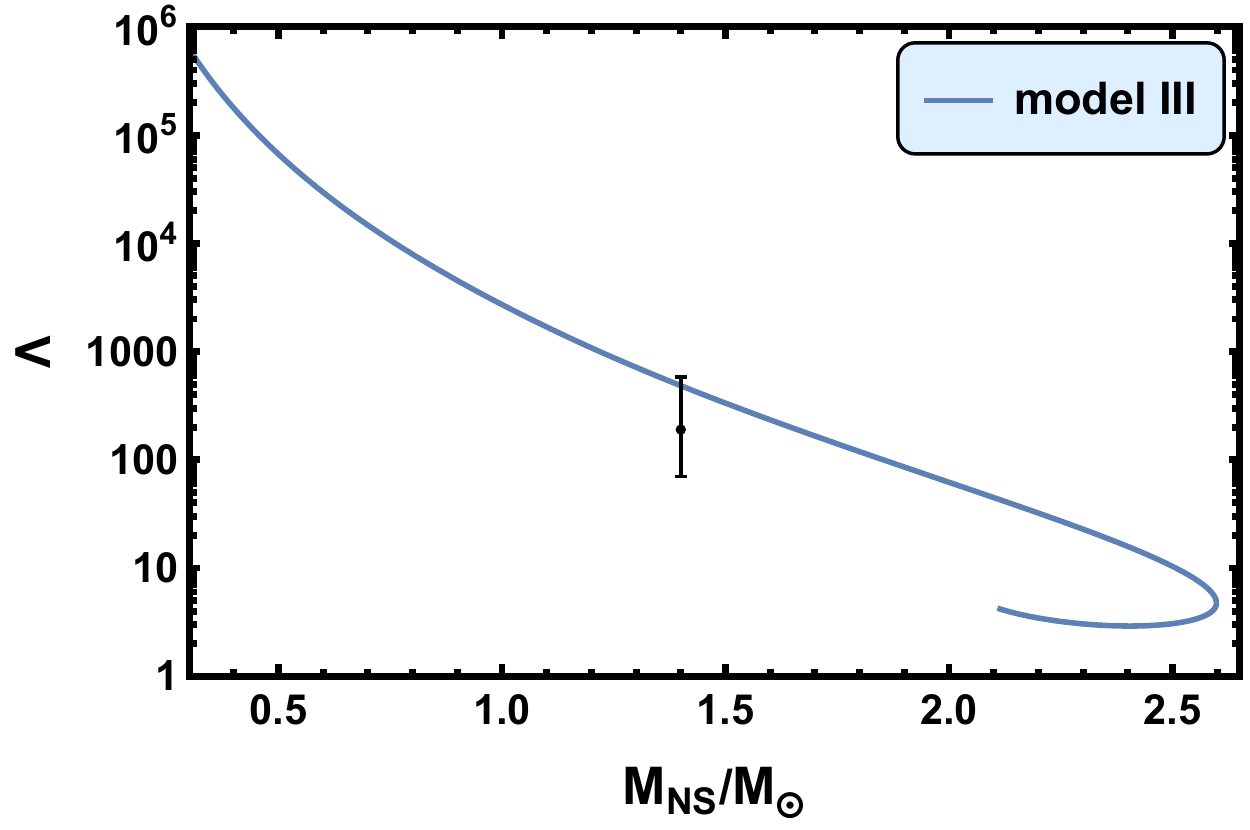}
  \caption{The tidal deformability as a function of the mass ratio between the neutron star and the sun calculated in model \RNum{1}, model \RNum{2}, and model \RNum{3}. The black dot with error bar is the tidal deformability of a $1.4 M_{\odot}$\hyp{}NS, estimated in Ref. \cite{LIGOScientific:2018cki}: $\Lambda_{1.4}=190_{-120}^{+390}$. In model \RNum{1}, our calculation result for the tidal deformability of a $1.4 M_{\odot}$\hyp{}NS is $\Lambda_{1.4}=484$. In model \RNum{2}, $\Lambda_{1.4}=472$. In model \RNum{3}, $\Lambda_{1.4}=484$. Our results of the tidal deformability in all three models are consistent with this experiment data.}
  \label{NS_propert_model_2}
\end{figure}

The $l=2$ Love number $k_2$ is \cite{Hinderer:2009ca}
\begin{align}
   & k_2=\frac{8 C^5}{5}\left(1-2 C\right)^2 \left[2+2 C (y-1)-y\right]
  \nonumber                                                                                       \\
   & \quad \times \bigg\{
  2C\left[6-3y+3C(5y-8)\right]
  \nonumber                                                                                       \\
   & \qquad +4 C^3 \left[13-11y+C\left(3y-2\right)+2 C^2 \left(1+y\right)\right]
  \nonumber                                                                                       \\
   & \qquad +3 \left(1-2 C\right)^2 \left[2-y+2 C \left(y-1\right)\right] \ln{\left(1-2 C\right)}
  \bigg\}^{-1},
  \label{l_2_Love_number_index_2}
\end{align}
where
\begin{align}
  y=\frac{r_B \beta(r_B)}{H(r_B)},
  \label{y}
\end{align}
and
\begin{align}
  C=\frac{G M}{r_B}
  \label{compactn}
\end{align}
is the compactness of the object.
From Eq. (\ref{l_2_Love_number_index_1}), we then define a dimensionless tidal deformability $\Lambda$ \cite{Kovensky:2021kzl}:
\begin{align}
  \Lambda=\frac{2}{3} \frac{k_2}{C^5}.
  \label{dimensionless_tidal_deformab}
\end{align}

The tidal deformability as a function of the mass ratio between the neutron star and the sun calculated in model \RNum{1}, model \RNum{2}, and model \RNum{3} are shown in Fig. \ref{NS_propert_model_2}. The black dot with error bar is the tidal deformability of a $1.4 M_{\odot}$\hyp{}NS, estimated in Ref. \cite{LIGOScientific:2018cki}: $\Lambda_{1.4}=190_{-120}^{+390}$. In model \RNum{1}, our calculation result for the tidal deformability of a $1.4 M_{\odot}$\hyp{}NS is $\Lambda_{1.4}=484$. In model \RNum{2}, $\Lambda_{1.4}=472$. In model \RNum{3}, $\Lambda_{1.4}=484$. Our results of the tidal deformability in all three models are consistent with this experiment data.

In Ref. \cite{LIGOScientific:2020zkf}, the authors gave that the radius of a canonical $1.4 M_{\odot}$ NS is $R_{1.4}=12.9_{-0.7}^{+0.8} \mathrm{km}$ and the tidal deformability is $\Lambda_{1.4}=616_{-158}^{+273}$ by assuming that the secondary of GW190814 proves to be a NS. Because the secondary mass of GW190814 is $m_2=2.59_{-0.09}^{+0.08} M_{\odot}$, that assumption requires that the max possible mass of the NS should be no less than $m_2$. Model \RNum{1} and model \RNum{3} satisfy this requirement. Actually, we find the results of $R_{1.4}$ and $\Lambda_{1.4}$ given by the three models are all consistent with those from Ref. \cite{LIGOScientific:2020zkf}.

Refs. \cite{Ma:2018qkg,Kanakis-Pegios:2020jnf,Kanakis-Pegios:2021uqc} also discussed the sound velocity and tidal deformability. For example, the authors in Ref. \cite{Ma:2018qkg} used the pseudo-conformal model \cite{Paeng:2017qvp,Ma:2018jze,Ma:2018xjw} and presented that $\Lambda_{1.4}=785$ for $n_{1/2}=2.0$, and $\Lambda_{1.4}=652$ for $n_{1/2}=3.0$ and $n_{1/2}=4.0$, where $n_{1/2}$ denoted the density at which the topology change takes place.

\section{Conclusion and discussion}
\label{sec_conclus_and_discus}

We construct the hybrid equations of state (EoSs) of the cold strong interaction matter using three holographic models in this work. In the low baryon number density $n_B$ range, the HLPS intermediate EoS is used to describe the nuclear matter. At the intermediate $n_B$ range, we use the holographic QCD to calculate the EoS. The EoS from the holographic QCD is connected with the HLPS EoS through the phase equilibrium condition. The key point is to describe the nuclear matter in the holographic frame, the EoS of which can be connected with the HLPS EoS through a second order phase transition. It is found that the branch that corresponds to the holographic nuclear matter is stable in the mass-radius relation of the NS.

The holographic action consists of two parts: the Einstein-Maxwell-dilaton (EMD) system for the background and the KKSS action for the matter part. The EMD system is solved to get the background metric which is used to calculate the EoS. For holographic nuclear matter, the thermodynamic contribution made by the KKSS action is also included to calculate the EoS. From the previous study in Ref. \cite{Chen:2019rez}, most probably the holographic nuclear matter is in the quarkyonic phase \cite{McLerran:2018hbz} in which the chiral symmetry is restored but the quarks and gluons are still confined. We will study the quarkyonic phase and its relationship with the NS in our holographic models in the future. There are already some articles \cite{Hata:2007mb,Ishii:2019gta,Jokela:2020piw,Jarvinen:2021jbd,Kovensky:2021ddl,Kovensky:2021kzl,Kovensky:2021wzu,Bartolini:2022rkl} to describe the baryons by introducing the Dirac-Born-Infeld (DBI) action and the Chern-Simons (CS) terms, where the solitons are regarded as baryons. However, the method to calculate the EoS of the holographic nuclear matter here is new and hadn't been used within the holographic QCD yet.

The hybrid EoSs we constructed are located in the ``allowed'' range constrained by interpolating the results of nuclear physics and the results of perturbative QCD with the constraint of the astrophysical observations. Then we calculate the properties of the neutron star, including the mass-radius relation and the tidal deformability.

We use three different holographic models here. In model \RNum{1}, the radius for the $1.4 M_{\odot}$\hyp{}NS is $12.320 \mathrm{km}$. The max possible radius for the NS is $12.833 \mathrm{km}$, which correspond a $2.32 M_{\odot}$\hyp{}NS. The max possible mass for the NS is $2.630 M_{\odot}$, which correspond a $12.559 \mathrm{km}$\hyp{}NS.
In model \RNum{2}, the radius for the $1.4 M_{\odot}$\hyp{}NS is $12.287 \mathrm{km}$. The max possible radius for the NS is $12.475 \mathrm{km}$, which correspond a $1.98 M_{\odot}$\hyp{}NS. The max possible mass for the NS is $2.466 M_{\odot}$, which correspond a $11.902 \mathrm{km}$\hyp{}NS.
In model \RNum{3}, the radius for the $1.4 M_{\odot}$\hyp{}NS is $12.319 \mathrm{km}$. The max possible radius for the NS is $12.795 \mathrm{km}$, which correspond a $2.27 M_{\odot}$\hyp{}NS. The max possible mass for the NS is $2.598 M_{\odot}$, which correspond a $12.24 \mathrm{km}$\hyp{}NS.

The branch that corresponds to the holographic quark matter is always unstable, and the holographic nuclear matter can produce the stable branch. These results indicate that even in the core of the NS, the matter is still in the confinement phase and the quark matter is not favored. This is consistent with the conclusion in Ref. \cite{Jokela:2021vwy}, in which the authors combined the input from the gauge\hyp{}gravity duality and the input from various $ab$-$initio$ methods for the nuclear matter at low density and analyzed families of hybrid equations of state of cold QCD matter. Their prediction is that all neutron stars are fully hadronic without quark matter cores.

In model \RNum{1}, our calculation result for the tidal deformability of a $1.4 M_{\odot}$\hyp{}NS is $\Lambda_{1.4}=484$. In model \RNum{2}, $\Lambda_{1.4}=472$. In model \RNum{3}, $\Lambda_{1.4}=484$. Our results of the tidal deformability in all three models are consistent with that estimated in Ref. \cite{LIGOScientific:2018cki}.
In Ref. \cite{LIGOScientific:2020zkf}, the authors gave that the radius of a canonical $1.4 M_{\odot}$ NS is $R_{1.4}=12.9_{-0.7}^{+0.8} \mathrm{km}$ and the tidal deformability is $\Lambda_{1.4}=616_{-158}^{+273}$ by assuming that the secondary of GW190814 proves to be a NS. Because the secondary mass of GW190814 is $m_2=2.59_{-0.09}^{+0.08} M_{\odot}$, that assumption requires that the max possible mass of the NS should be no less than $m_2$. Model \RNum{1} and model \RNum{3} satisfy this requirement. Actually, we find the results of $R_{1.4}$ and $\Lambda_{1.4}$ given by the three models are all consistent with those from Ref. \cite{LIGOScientific:2020zkf}.

For future studies, we will investigate the relation between the chiral phase transition and confinement-deconfinement phase transition, and figure out whether the stable holographic nuclear matter is in the quarkyonic phase. Currently, the EMD system and the KKSS action are solved separately, which means we use the ``probe approximation'' here. The KKSS action is regarded as a probe and its back\hyp{}reaction to the background EMD system is neglected. We will investigate this holographic model beyond the probe approximation and solve it consistently in the future. Also, as we stated in the introduction, the ``allowed'' range of the EoS we used here is derived by the ``direct interpolation approach'' \cite{Annala:2017llu,Annala:2021gom}. However the ``causality and thermodynamic stability constraint approach'' proposed in Refs. \cite{Komoltsev:2021jzg,Gorda:2022jvk} will give more strict constraints on the EoS of cold QCD matter. We will consider these constraints in future works.



\begin{acknowledgments}
  We thank Minghua Wei and Yidian Chen for helpful discussions. This work is supported in part by the National Natural Science Foundation of China (NSFC) Grant Nos. 11725523 and 11735007, the Strategic Priority Research Program of Chinese Academy of Sciences under Grant Nos XDB34030000 and XDPB15, the start-up funding from University of Chinese Academy of Sciences(UCAS), and the Fundamental Research Funds for the Central Universities.
\end{acknowledgments}


\bibliographystyle{unsrt}
\bibliography{references.bib}

\end{document}